\documentclass[
 reprint,
 amsmath,amssymb,
 aps,
]{revtex4-2}
\usepackage[T1]{fontenc}
\usepackage{graphicx}
\usepackage{dcolumn}
\usepackage{bm}
\usepackage{amsmath}
\usepackage{lmodern}
\usepackage{mathtools}
\usepackage{caption}
\usepackage{siunitx}
\usepackage{gensymb}
\usepackage{caption}
\usepackage{subcaption}
\usepackage{lipsum}
\usepackage{float} 
\usepackage[colorlinks,citecolor=black,linkcolor=black,urlcolor=black]{hyperref}
\usepackage[normalem]{ulem}

\begin{document}

\title{Molecular Beam Epitaxy of a 2D material nearly lattice matched to a 3D substrate: $NiTe_{2}$ on $GaAs$}
\author{B.~Seredyński$^1$}
\email{Bartlomiej.Seredynski@fuw.edu.pl}
\author{Z.~Ogorzałek$^1$}
\author{W.~Zajkowska$^2$}
\author{R.~Bożek$^1$}
\author{M.~Tokarczyk$^1$}
\author{J.~Suffczyński$^1$}
\author{S.~Kret$^2$}
\author{J.~Sadowski$^{1,2,3}$}
\author{M.~Gryglas-Borysiewicz$^1$}
\author{W.~Pacuski$^1$}
\email{Wojciech.Pacuski@fuw.edu.pl}

\affiliation{$^1$Institute of Experimental Physics, Faculty of Physics, \\University of Warsaw, ul. Pasteura 5, PL-02-093 Warsaw, Poland}
\affiliation{$^2$Institute of Physics, Polish Academy of Sciences,\\ Aleja Lotnik\'ow 32/46, PL-02-668 Warsaw, Poland}
\affiliation{$^3$Department of Physics and Electrical Engineering, Linnaeus University, SE-391 82
Kalmar, Sweden}


\begin{abstract}
    The lattice mismatch between interesting 2D materials and commonly available 3D substrates is one of the obstacles in the epitaxial growth of monolithic 2D/3D heterostructures, but a number of 2D materials have not yet been considered for epitaxy. Here we present the first molecular beam epitaxy growth of NiTe$_{2}$ 2D transition metal dichalcogenide. Importantly, the growth is realized on a nearly lattice matched GaAs(111)B substrate. Structural properties of the grown layers are investigated by electron diffraction, X-ray diffraction, and scanning tunnelling microscopy. Surface coverage and atomic scale order is evidenced by images obtained with atomic force, scanning electron, and transmission electron microscopy. Basic transport properties were measured confirming that NiTe$_{2}$ layers are metallic, with the Hall concentration of $10^{20}$~cm$^{-3}$ to $10^{23}$~cm$^{-3}$, depending on the growth conditions.
\end{abstract}
\maketitle

\section{\label{sec:intro}Introduction} 
	The extensive research activity in the field of 2D materials does not weaken since the advent of graphene and other layered materials \cite{Graphene, Layered_BP, Layered_Halides_1, Layered_Halides_2}. Especially the class of the layered transition-metal dichalcogenides (TMDs) plays a great role in the field \cite{2D_TMDC_book, TMDC_general}. Because of their unique properties associated with the fascinating electronic structure TMDs are potential candidates for applications in various fields {}
	such as optics \cite{TMDC_laser_1,TMDC_laser_2}, nanoelectronics \cite{Nanoelectronics_1,Nanoelectronics_2}, energy storage \cite{EnergyStorage_1,EnergyStorage_2} or sensing \cite{Sensing}. Recently they also gained interest because of topological properties. For example PtTe$_{2}$ belongs to type-II Dirac semimetals \cite{PtTe2_Dirac}, while WTe$_{2}$ and MoTe$_{2}$ (in one of its polytypes) to a type-II Weyl semimetals \cite{MoTe2_Weyl}. NiTe$_{2}$ was recently identified as a member of the first group (i.e. type-II Dirac semimetal). Recent reports show electronic structure calculations \cite{ACS_type2} or angular resolved photoelectron spectroscopy results \cite{exfoliated} proving topological nature of NiTe$_{2}$. Accordingly, NiTe$_{2}$ topological properties helps to achieve high frequency operation devices in a terahertz technology applications \cite{NiTe2_THz_1, NiTe2_THz_2}.
	
	The best electronic and optical properties of 2D materials are typically obtained for samples exfoliated from bulk crystals \cite{Exf_MoTe2, Golasa_MoS2}. The exfoliation technique is however limited by a lack of scalability and reproducibility. Therefore, many efforts are focused on the epitaxial growth of 2D materials demonstrating that the properties of as grown layers are comparable with exfoliated ones \cite{MoSe2_nasze} and the surface area reaches commercial substrate wafer size. Despite the problem of the lattice mismatch between available substrates and the TMD epi-layers there are numerous reports of successful epitaxial growth of various 2D materials using chemical methods, such as Chemical Vapour Deposition (CVD) \cite{CVD_2D,CVD_gold}, and physical methods, such as Molecular Beam Epitaxy (MBE) \cite{Large_area_MoSe2, Furdyna_przeglad, MoTe2_nasze, MoSe2_nasze, MBE_TMDC_tellurides, MBE_TMDC_selenides, MoSe2_orientation}.
    The crystals of NiTe$_{2}$ have been produced so far only by the chemical ones including several bulk growths realizations \cite{Bulk_1, Bulk_2, Bulk_3, Bulk_4, Bulk_5}. 
    Thin films of NiTe$_{2}$ were successfully grown by CVD on SiO$_{2}$\cite{CVD_JACS} where the precise thickness control was reported for lateral dimension up to hundreds of micrometers. Other CVD growths of NiTe$_{2}$ thin layers presented in literature include growth on MoS$_{2}$ \cite{NiTe2_na_MoS2} and mica \cite{CVD_mica} substrates which lattice constants substantially differ from that of NiTe$_{2}$. 
	
    In this paper we present MBE growth of NiTe$_{2}$ on a nearly lattice matched substrate 3D GaAs(111). A number of experimental techniques are used to characterise the obtained layers. To our knowledge this is the first report on NiTe$_{2}$ grown by the molecular beam epitaxy.


        \begin{figure}
            \centering
            \includegraphics[width=\linewidth]{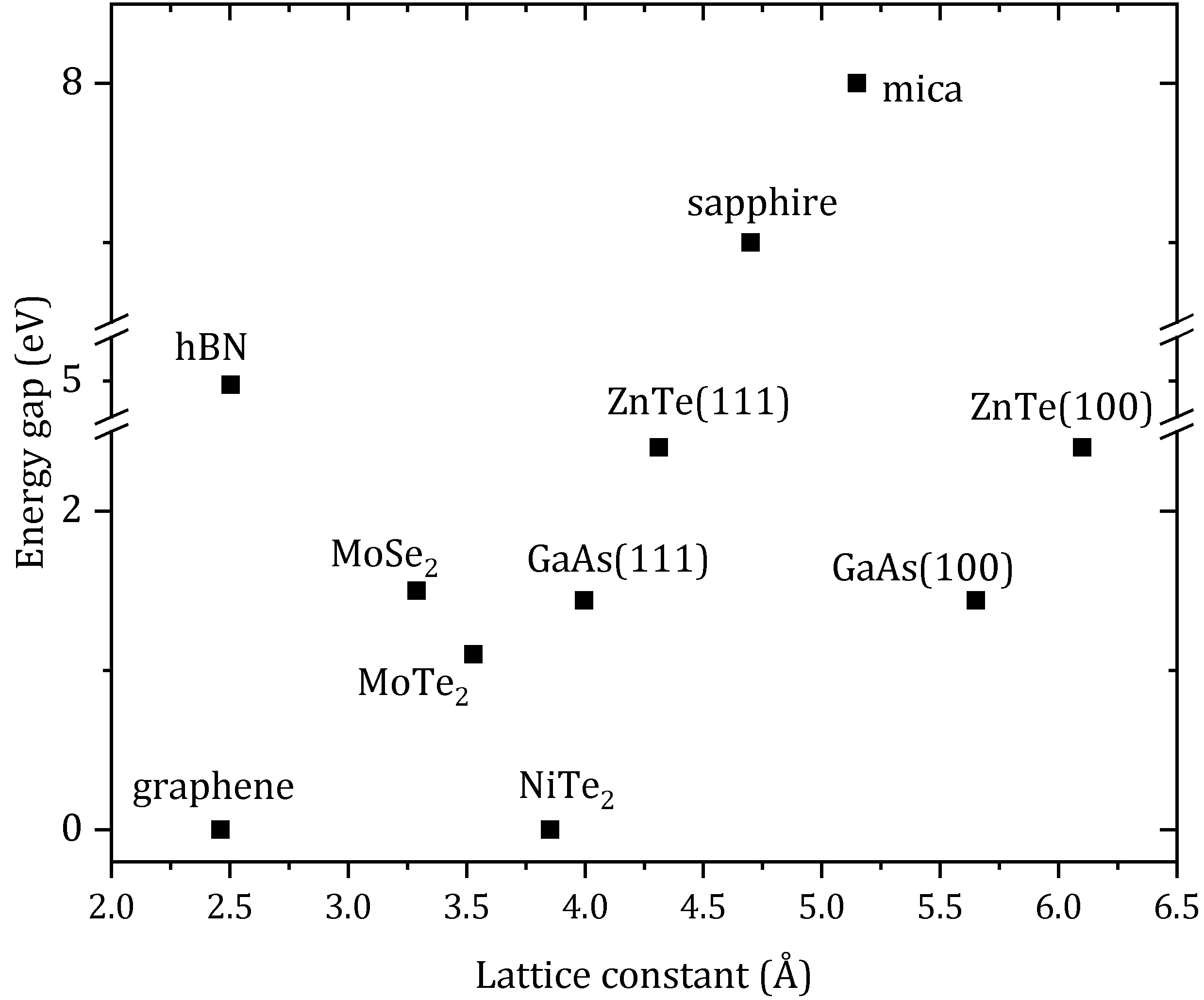}
            \caption{Plot of common 2D materials and 2D/3D substrates with energy gaps and lattice parameters. For crystals in zinc blend structure, lattice parameters within (111) plane are shown because the symmetry of this plane fits to 2D materials. Due to relatively small lattice mismatch i.e. 3.6 \%, GaAs(111) is a promising substrate for the epitaxy of NiTe$_{2}$.}
            \label{im: Lattices}
        \end{figure}    


\section{\label{sec:sample}Sample growth}
	The growth was performed in the MBE chamber dedicated to II-VI semiconductors. Working pressure in the growth chamber was maintained bellow $10^{-9}$ mbar. Both nickel and tellurium were evaporated from conventional effusion cells: high temperature cell with aluminium oxide crucible for Ni and low temperature, dual filament cell with pBN crucible for Te. The growth was performed in the excess of tellurium molecular flux (with the ratio of Te/Ni fluxes of about 40:1). The samples are grown on a half of 2 in semi-insulating epi-ready GaAs(111)B substrate. The choice of this substrate is related to the fact, that the GaAs(111)B surface shares the trigonal symmetry with NiTe$_{2}$. Moreover the surface lattice constant of GaAs(111), $a=3.997$ \AA{} \cite{GaAs_lattice} matches very well that of NiTe$_{2}$, $a=3.855$ \AA{} \cite{ACS_type2}. This makes GaAs(111) almost a perfect substrate for NiTe$_{2}$ growth (see Fig.~\ref{im: Lattices}).
	Prior to the growth the substrate was $in-situ$ annealed above 600~{\degree}C to remove native oxides. Then the substrate was cooled down to 330~{\degree}C growth temperature. For the transport measurements we also discuss lower growth temperature of 280~{\degree}C and a growth with a temperature step i.e. half of the growth was performed at 330~{\degree}C and the second half at 280~{\degree}C. Both Ni and Te shutters were opened simultaneously. The samples were not annealed after the growth.


\section{\label{sec:experiment}Experimental results}
		Below we discuss the results giving insight to the properties of NiTe$_{2}$ epilayers. Wafer scale high crystalline quality of NiTe$_{2}$ grown on GaAs(111)B is proven by Reflection High Energy Electron Diffraction (RHEED) images and X-ray Diffraction (XRD) 2$\theta / \omega$ scans. Surface coverage and roughness are evaluated with Atomic Force Microscopy (AFM) and Scanning Electron Microscopy (SEM). Atomic-scale crystal alignment is revealed by Scanning Tunnelling Microscopy (STM) and Transmission Electron Microscopy (TEM) images. Resistivity as a function temperature is discussed. Room temperature Raman spectroscopy results are also presented.


	\subsection{Reflection High Energy Electron Diffraction}	
\begin{figure}
            \centering
            \includegraphics[width=0.7\linewidth]{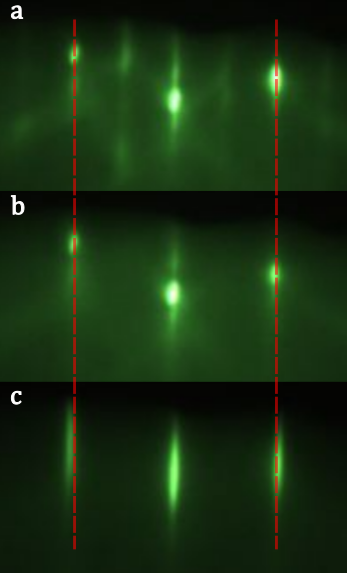}
            \caption{RHEED patterns registered for different stages of NiTe$_{2}$ growth. a) deoxidized GaAs(111)B substrate before the growth taken along the $[1\overline{1}0]$ direction, b) 1~minute of the NiTe$_{2}$ growth and c) 50 nm thick NiTe$_{2}$ taken along $[\overline{1}\overline{1}20]$ direction. Red dashed lines are parallel to observed streaks and serve as guides for the eye to evidence similar lattice parameters of both materials.}
            \label{im:RHEED}
        \end{figure}

    	Figure \ref{im:RHEED} presents evolution of electron diffraction pattern during the sample growth. First, as shown in Fig. \ref{im:RHEED}a the partially spotty RHEED pattern related to the GaAs(111)B surface emerges. The surface is partly rough due to just finished procedure of thermal deoxidation. However it is smooth enough to exhibit clear (2~$\times$~2) reconstruction \cite{GaAs111_reconstruction} observed as two additional lines between three main streaks. Figure \ref{im:RHEED}b shows RHEED image obtained just one minute after the start of the of NiTe$_{2}$ growth (after deposition of about  1 nm of material). GaAs(111)B surface reconstruction is not observed any more and the image is more spotty, suggesting island growth mode. However, the next few minutes of the growth the flattening of the surface which is observed as a streaky pattern with sharp lines, as shown on the RHEED image displayed in Fig. \ref{im:RHEED}c. The distances between two main streaks of GaAs and those of NiTe$_{2}$ in RHEED images are very similar which proves very close in-plane lattice parameters of both materials, and confirms motivation of our work (growth on almost lattice matched substrate). Knowing the value of the GaAs lattice constant one can determine the in-plane lattice constant of the as grown material, NiTe$_{2}$ here. Comparing distances between satellite streaks the resulted lattice constant of NiTe$_{2}$ is estimated to (3.89 $\pm$ 0.04) \AA{} which is between the literature value of bulk NiTe$_{2}$ (3.855 \AA{}) \cite{ACS_type2} and GaAs(111) (3.997 \AA{}) \cite{GaAs_lattice}. This indicates only partial strain relaxation.
    	Observation of streaky RHEED pattern during the growth of NiTe$_{2}$  is an evidence of truly epitaxial growth with crystal axes of the epi-layer determined by the crystal axes of the substrate.

	\subsection{X-Ray Diffraction}	
\begin{figure}
            \centering
            \includegraphics[width=\linewidth]{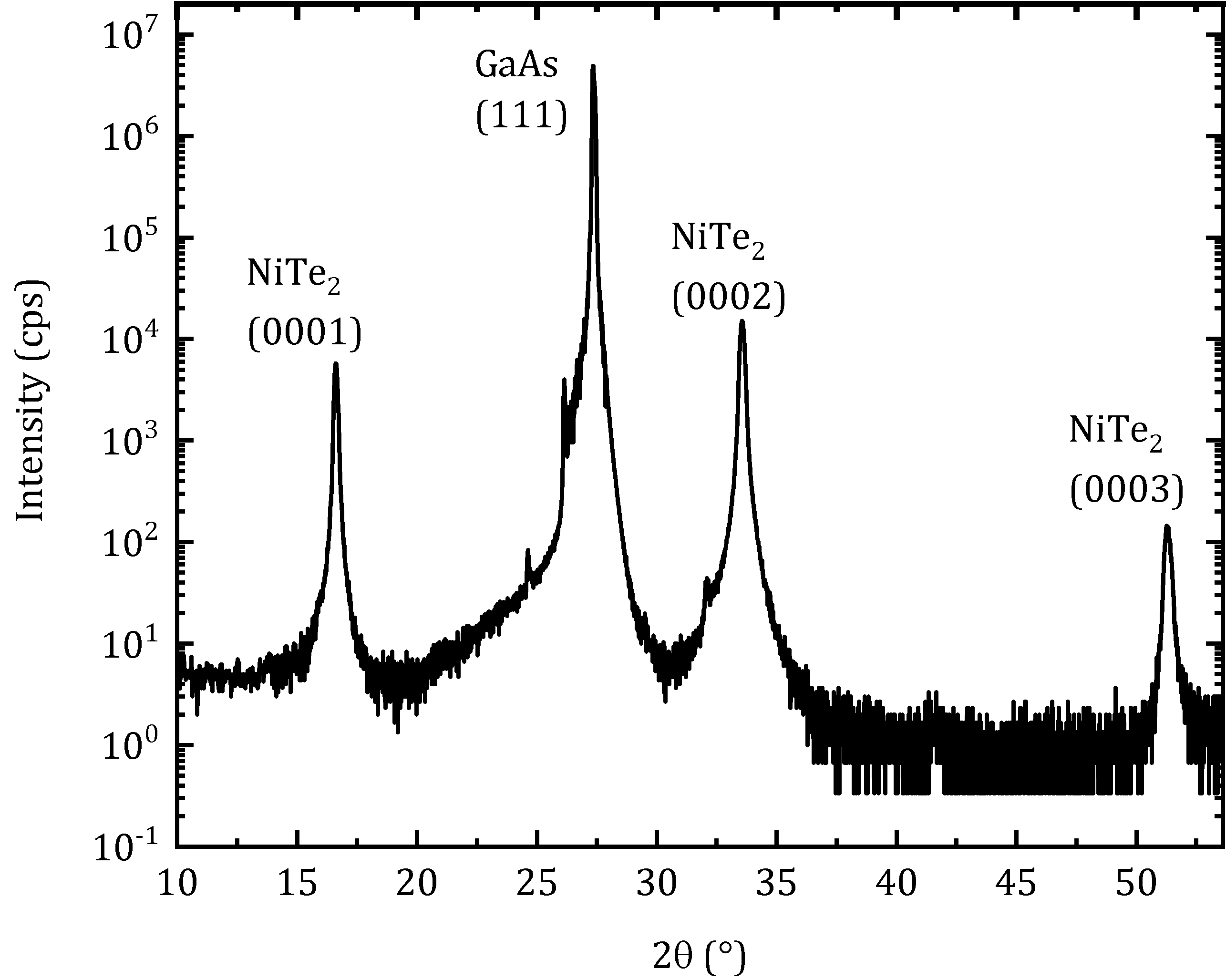}
            \caption{2$\theta / \omega$ scan of NiTe$_{2}$. Consecutive (0001), (0002) and (0003) peaks indicate good epitaxial orientation of the as grown NiTe$_{2}$ layers. The low-intensity GaAs(111) peak is due to not fully monochromatic X-ray beam.}
            \label{fig:XRD}
        \end{figure}

        Figure \ref{fig:XRD} presents 2$\theta / \omega$ scan of NiTe$_{2}$ epitaxially grown on GaAs(111). The presence of NiTe$_{2}$ (0001), (0002) and (0003) peaks indicates flatness of the layers and its [001] orientation. The positions of observed peaks agree well with both databases for NiTe2$_{2}$: International Centre for Diffraction Data card no. 00-008-0004 \cite{baza_XRD} and no. 01-089-2642 \cite{baza_XRD2} The out-of-plane lattice parameter obtained equals (5.34~$\pm$~0.05)~\AA{} which agrees well with the literature values \cite{ACS_type2, XRD_stala_c}.  Peak around 27$\degree$ originates from the GaAs substrate. All observed peaks are identified suggesting absence of unintentionally introduced secondary phases.

	\subsection{Scanning Tunnelling Microscopy}	
\begin{figure}
    		\centering
    		\includegraphics[width=0.82\linewidth]{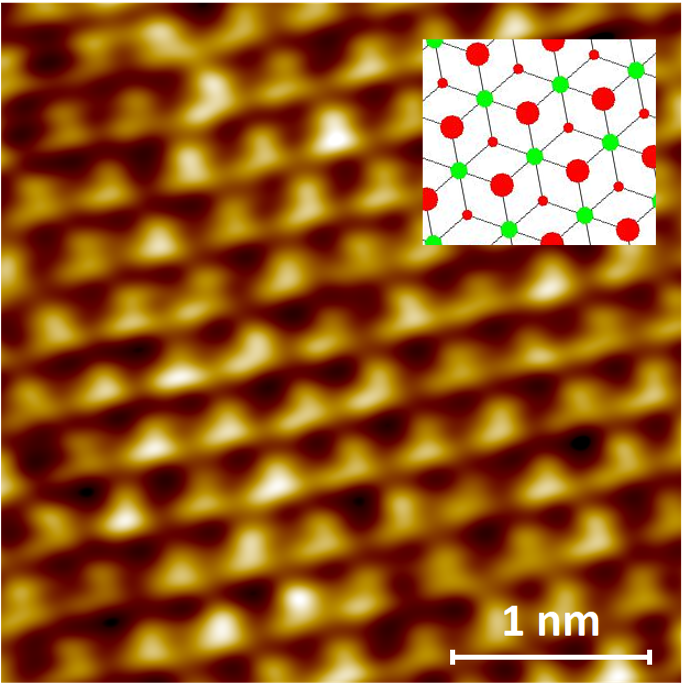}
    		\caption{STM image of NiTe$_{2}$ layer measured with sample bias V$_b$ = +10 mV and I$_t$ = 1.5~nA. The inset shows the stick-and-ball model of a NiTe$_{2}$ monolayer with green and red balls representing nickel and tellurium atoms respectively. The diameter of the ball represents depth of the atom. The averaged in-plane lattice parameter is equal to 0.38 nm.}
    	    \label{im:STM}
    	\end{figure}

	STM measurements of epitaxial NiTe$_{2}$ surface were performed in ultra high vacuum and at room temperature. Example STM image with atomic resolution is presented in Fig. \ref{im:STM}. The surface is atomically flat with a characteristic pattern. With the distance between consecutive maxima we estimate the in-plane lattice constant of NiTe$_{2}$ to be around 3.8 \AA{}~which agrees very well with the reported values \cite{ACS_type2} and the value obtained in this work from the RHEED pattern.

    	
    \subsection{Atomic Force Microscopy and Scanning Electron Microscopy}
\begin{figure*}
            \centering 
            \includegraphics[width=1\linewidth]{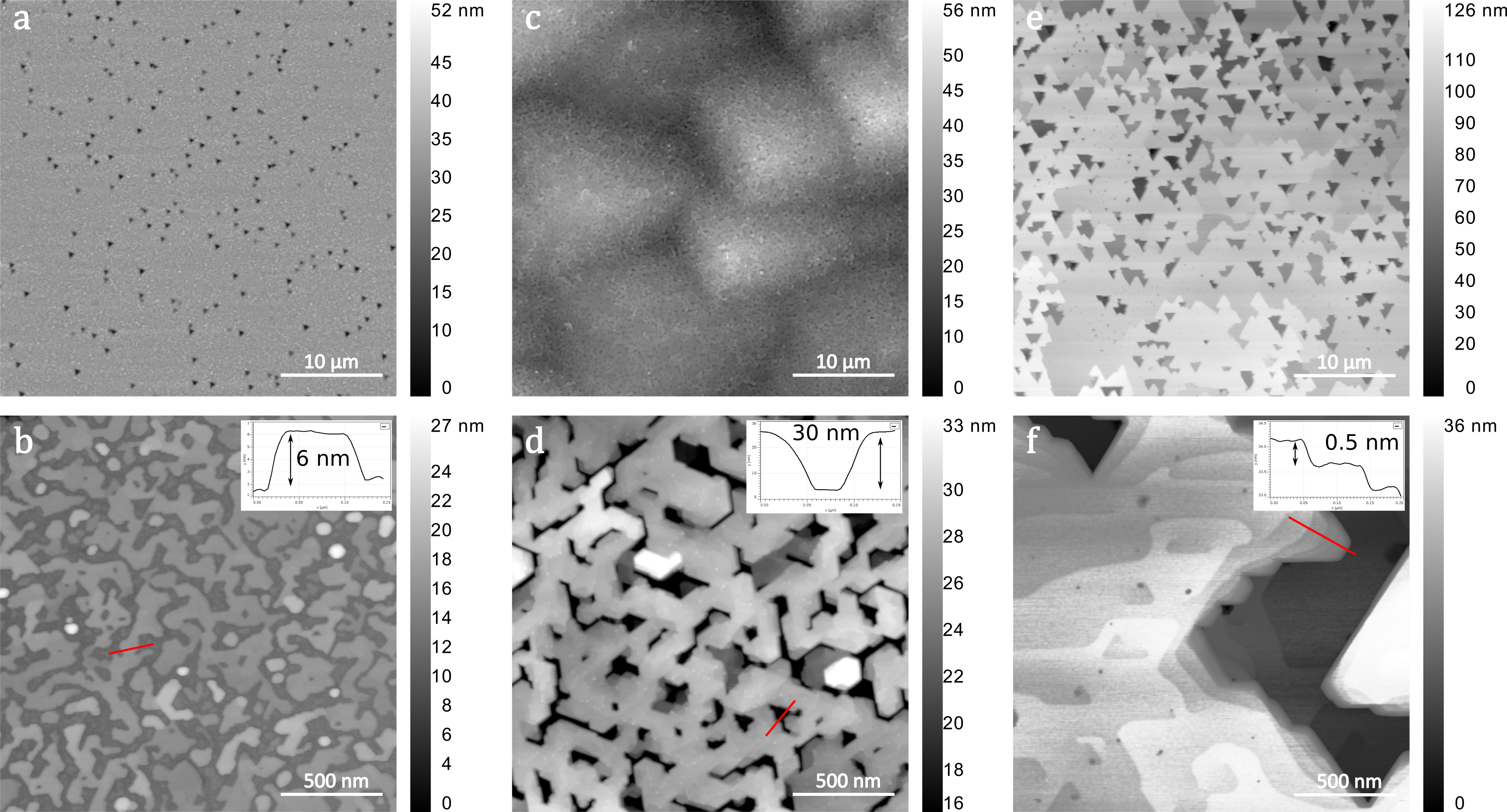} 
            \caption{AFM images of NiTe$_{2}$ with various average thickness: a, b) 6~nm thick, c, d) 30~nm thick e,f) 500~nm thick. Scanned area is  $20~\mu $m $\times$  $20~\mu $m large for the top row and $2~\mu $m $\times$  $2~\mu $m large for the bottom row of the figure. Vertical scales in figures d) and f) are adjusted to reveal details  of NiTe$_{2}$ surface. In the insets of figures b), d) and f) cross-sections marked by red lines on corresponding images are embedded. For b) and d) the marked height corresponds to an average sample thickness, for f) the value of 0.5 nm corresponds to a single NiTe2 atomic step.}
            \label{im:AFM}
        \end{figure*}

\begin{figure}
            \centering 
            \includegraphics[width=0.82\linewidth]{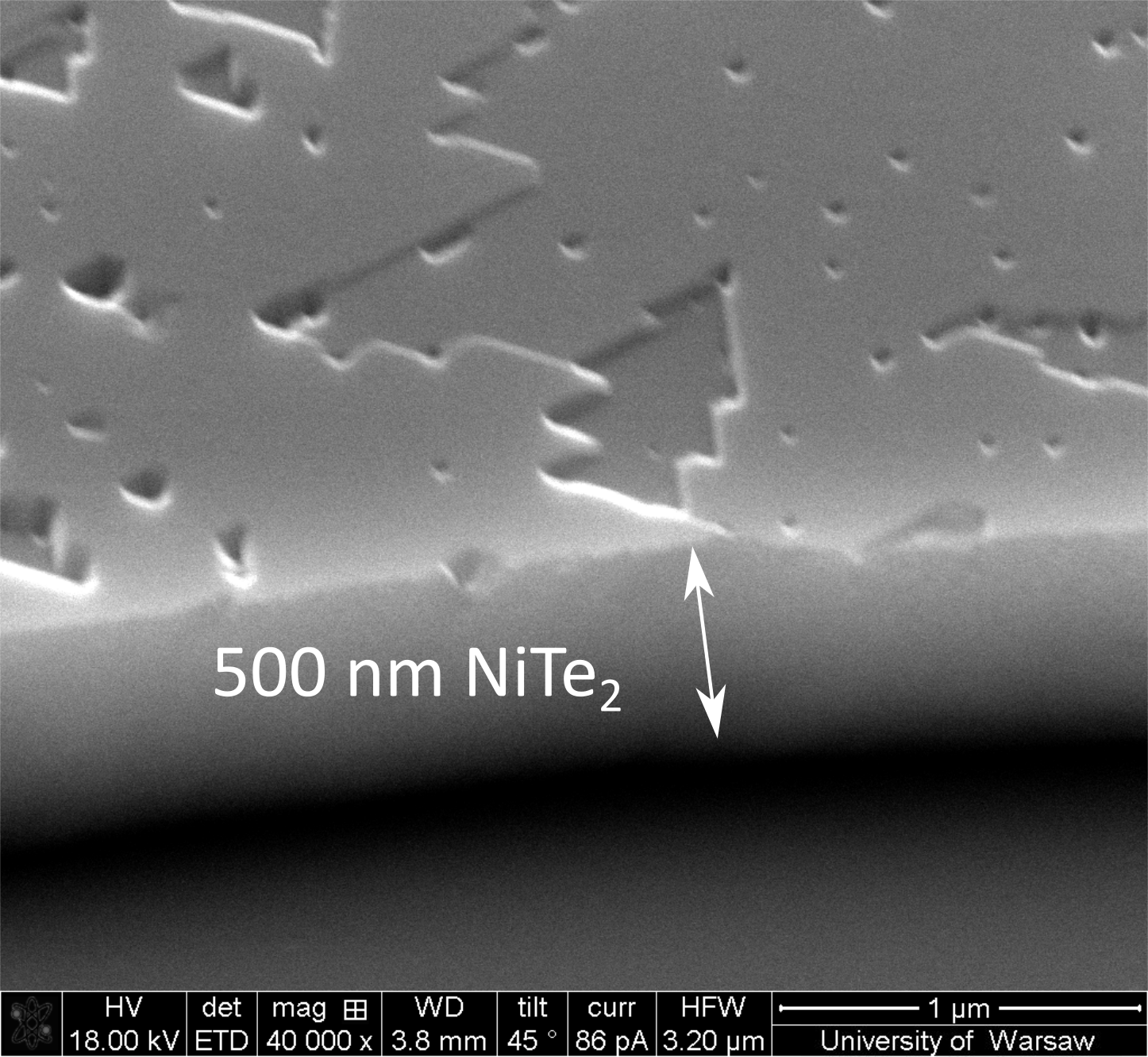}
            \caption{SEM image of 500 nm thick NiTe$_{2}$ showing the edge of the sample and highlighting its continuity.}
            \label{im:SEM}
        \end{figure}

    	AFM images of 20 $\mu $m $\times$ 20 $\mu $m and 2 $\mu $m $\times$  2 $\mu $m area of three samples with different thicknesses are shown in Fig. \ref{im:AFM}. The images reveal temporal evolution of the growth mode. In particular Fig. \ref{im:AFM} b a shows that for 6~nm  thick NiTe$_{2}$ there are gaps in the surface i.e. NiTe$_{2}$ does not cover the whole substrate surface. This indicates the island growth mode rather than the planar layer-by-layer growth, in agreement with the conclusions based on the observation of spotty RHEED patterns in the beginning of the growth. For the thicker, 30~nm sample (Fig. \ref{im:AFM}d)  gaps between the islands start to decrease but still leaving some pits reaching the substrate. 
    	Finally, for the 500~nm thick sample the pits have depths of around 100~nm but no holes reaching the substrate appear. Here, in contrast to two previous samples, almost atomically flat surface is obtained as shown in Fig. f). The inset presents the cross-section marked with the red line on the image. The 0.5~nm height corresponds to a single atomic step of NiTe$_{2}$.
    	Figure~\ref{im:SEM} presents SEM image of the edge of the 500 nm thick sample showing its continuous character and confirming its thickness.

	\subsection{Transmission Electron Microscopy}	

\begin{figure}
            \centering 
            \includegraphics[width=1\linewidth]{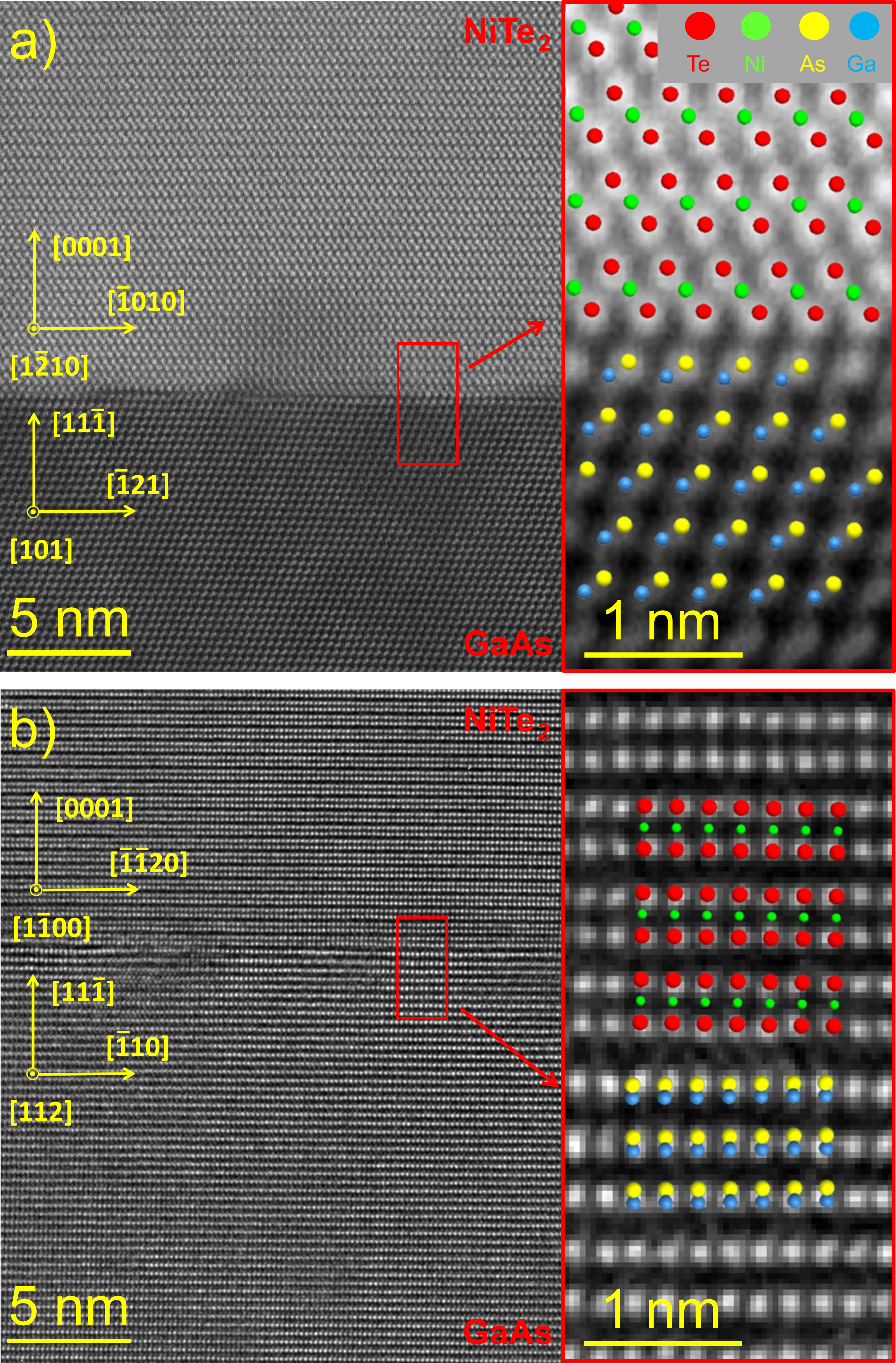}
            \caption{HR-STEM images of 15 nm thick NiTe$_{2}$ on GaAs. a) view towards GaAs [101] direction, b) view towards GaAs [112] direction. The interface between GaAs and NiTe$_{2}$ is smooth and sharp. Next layers of NiTe$_{2}$ lay on each other. Zoomed part in frame shows detail of interface region – coherent part.}

    		\label{im:TEM}
    	\end{figure}

The high-resolution scanning transmission electron (HR-STEM) microscope images of the cross-section of the 15 nm thick NiTe$_{2}$ are shown in Fig. \ref{im:TEM}. Figure \ref{im:TEM}a presents the view towards the [101] crystallographic direction and Fig. \ref{im:TEM}b towards [112] direction. In both figures one can observe sharp interface between GaAs (bottom of each image) and NiTe$_{2}$ (top of each image). In this projection Ga and As atoms cannot be separated or distinguished and appear as single dots, however position of Te atoms can easily be identified. Ni atoms are not well visible between heavy Te atoms. The alignment of the consecutive NiTe$_{2}$ layers is nearly perfect. They lay parallel on each other. This proves good growth conditions partly enabled by the lattice constant match. Insets in the red frames present zoomed part of the interface regions of the images with specific atoms marked with the corresponding colors. There are no defects present in  NiTe$_{2}$ layer within tens of nanometres. The epitaxial relations are as follows: $[\overline{1}010]$ NiTe$_{2}$ $\parallel$ $[1\overline{2}\overline{1}]$GaAs, $[\overline{1}\overline{1}20]$ NiTe$_{2}$ $\parallel$ $[1\overline{1}0]$ GaAs. In the Figure \ref{im:TEM}b the local $[1\overline{1}0]$ planes of GaAs are continued as $[\overline{1}\overline{1}20]$ planes of NiTe$_{2}$ without any visible shift. However, also the areas with lost coherency appear periodically.
The images suggest that in the interface the arsenic terminated surface connects directly with tellurium atoms of NiTe$_{2}$  in areas with good coherency. This suggests that there is some kind of additional interaction between As and Te in the coherent areas. However, based on our data, the atomic arrangements in the incoherent areas cannot be clearly reconstructed.


	\subsection{Transport measurements}

\begin{figure*}
    	\centering
    	\includegraphics[width=1\linewidth]{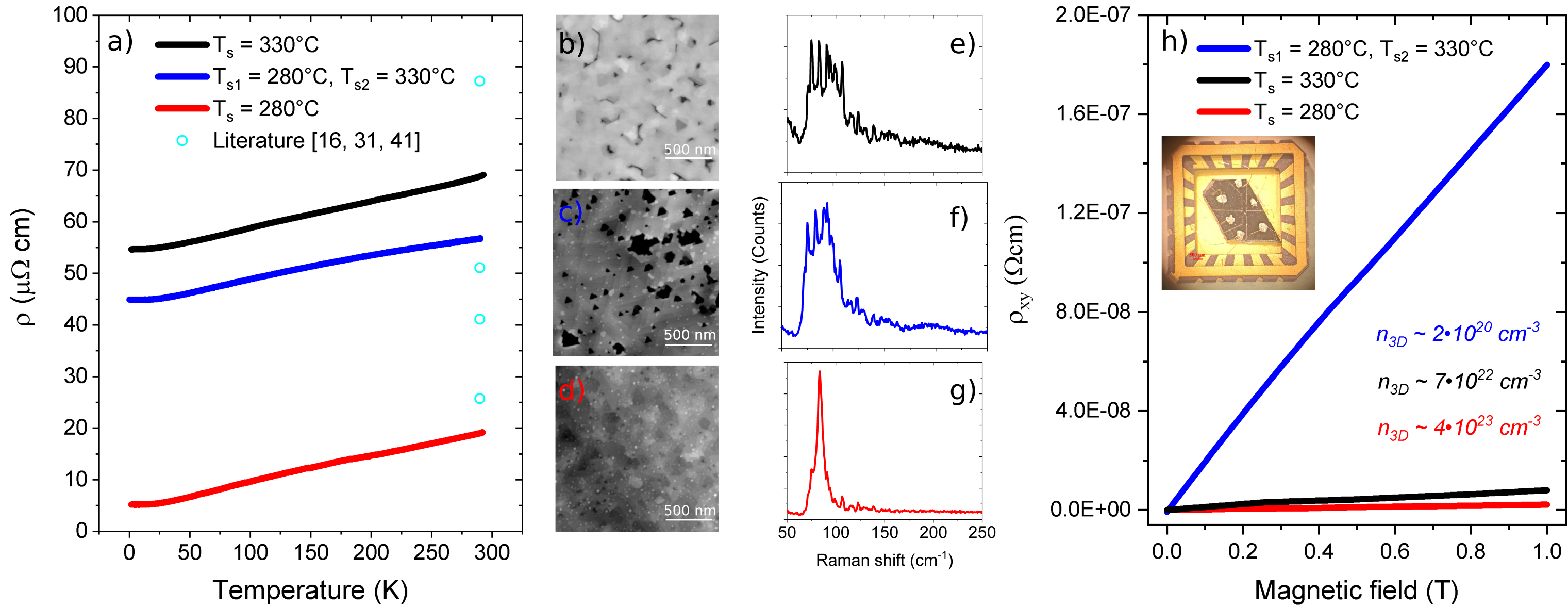}
    	\caption{a) The resistivity dependence on temperature for MBE-grown NiTe$_{2}$ layers and room temperature resistivity values for literature results of samples grown with other techniques \cite{Bulk_3, PRB2020, ACS_type2}. AFM images of samples grown in (b) 280~\degree~C, (c) 330~\degree~C and (d) at 280~\degree~C~/~330~\degree~C. (e-g) corresponding Raman spectra measured with 532 nm excitation laser. h) Transverse resistivities at room temperature for  samples grown at various temperatures. The inset presents the optical image of the Hall device. }
    		\label{im: Transport}
    \end{figure*}

     Large scale, 2-inch samples, were cut into smaller pieces and samples in the hallbar configuration were prepared. The width of the current path was 250 $\mu$m and the distance between voltage probes was about 500~$\mu$m. Resistivity versus temperature and Hall effect at room temperature were measured to provide information on basic transport properties of the samples. Figure \ref{im: Transport}a presents a measurement of resistivity versus temperature for 3 samples, grown at various temperatures. Red and black curves present results for the samples grown for 60 and 20 minutes at 280 \degree C and 330 \degree C respectively. Their average thicknesses are 50 nm and 15 nm accordingly. Results presented by the blue curve were obtained for the sample grown for 60 minutes, with the first half of the growth time performed at 280~\degree C and then at 330~\degree C. The average sample thickness is about 50 nm. The values of the room temperature resistivities correspond well to the values reported in the literature \cite{ACS_type2, Bulk_3, PRB2020} for samples grown by other techniques, and are presented by the open cyan points on the Fig. \ref{im: Transport}a. The resistance decreases with decreasing temperature, confirming the metallic character of the samples. The resistivity drop stays relatively small, which most probably is related to the large scale structural disorder not captured by the local STEM images. The topographies of the corresponding samples are shown on the AFM images in Figures \ref{im: Transport} b-d. The highest temperature induced drop in the relative resistivity correlates with the sharp Raman spectrum peak around $85$~cm$^{-1}$ (Fig. \ref{im: Transport}g). The position of the peak is in agreement with previous reports on Raman scattering in NiTe$_{2}$. \cite{SERS_Raman, exfoliated, NiTe2_na_MoS2}
     Figure \ref{im: Transport}h presents Hall resistivites at room temperature for NiTe$_{2}$ layers grown at various growth temperatures (T$_{s}$). The room temperature Hall effect measurements revealed that samples have linear Hall resistivities with n-type conductivity. The Hall concentrations vary from $10^{20}$~cm$^{-3}$ to $10^{23}$~cm$^{-3}$ for the substrate temperature T$_s$ from 280~\degree C to 330~\degree C. The control of the growth conditions permits to vary the concentration by two orders of magnitude, paving the way to control the position of the Fermi level for Dirac fermion studies.  

\section{\label{sec:conclusions}Conclusions}
	Monocrystalline layers of NiTe$_{2}$ II-type Dirac semimetal were grown for the first time using molecular beam epitaxy method. Material exhibits good crystalline quality thanks to the growth on nearly lattice matched GaAs(111)B substrate. The influence of the MBE growth conditions and layer thicknesses on the structural and morphological properties of NiTe$_{2}$ layers were studied. Even though the cross-sectional transmission electron microscopy images reveal nearly perfect NiTe$_{2}$-GaAs(111)B interfaces, the thin NiTe$_{2}$ layers are discontinuous, with surface gaps reaching the substrate. Only the thickest (500 nm) NiTe$_{2}$ layer investigated by us is free from deep holes.  The best quality NiTe$_{2}$ layers were grown at relatively low substrate temperature (280 C). 
	The samples are metallic with the Hall concentrations staying in the range $10^{20} cm^{-3}$ to $10^{23} cm^{-3}$. This spread of the Hall concentrations may reveal Fermi level shift but most probably indicates multiband conductivity in this material. This aspect needs further studies. 
	
\section{\label{sec:methods}Methods}	

X-ray diffraction measurements: 
All XRD measurements were obtained with Panalytical X'pert diffractometer equipped with a parallel beam Bragg X-ray mirror in front of the Cu X-ray tube (very weak Cu K $\beta$  and W L$\alpha$ lines may be present). The detector optics were dedicated to thin film analysis (Soller slits). Crystal structure information of NiTe$_{2}$  was obtained from 2 $\theta$/$\omega$ scans that were performed using a small angular step size (0.005~deg).

Scanning Tunnelling Microscopy:
STM measurements were performed at room temperature in UHV using an Omicron VT XA microscope. The sample was transferred to the chamber without any protective cap. In UHV it was annealed in the temperature  around 150~{\degree}C.

Atomic Force Microscopy:
AFM measurements were performed in a tapping mode using a Digital Instruments MultiMode AFM with a~Nanoscope IIIa controller.

Transmission Electron Microscope:
The morphology and structure of NiTe$_{2}$ on GaAs(111)B have been investigated using a FEI Titan 80-300 transmission electron microscope operating at 300 kV, equipped with an image corrector. The electron transparent cross-sections of heterostructure in two different orientations rotated by 30 \degree have been cut using a focused gallium ion beam in Helios Nanolab 600 focused ion beam (FIB). Prior to the ion cutting, the surface of the heterostructure was protected by an electron deposited Pt–C composite from a metalorganic source. The scanning transmission electron microscopy-high angle annular dark field detector – (STEM-HAADF) images have been acquired at the scattering angle range between 80 and 200 mrad camera length, with a converged semi angle of 9.5 mrad of the incident beam. The upper scattering angles for electron were limited, in our case,  by cut-off introduced by image corrector.

Electrical Measurements:
The samples were cleaved and cut into rectangles of about 4 mm x 2 mm and the silver contacts were manually applied. Neither chemical treatments nor solvents were applied in the procedure to avoid any contamination of the NiTe$_{2}$. DC and AC transport measurements were performed in a cryostat in a magnetic field up to 1 T with Variable Temperature Insert (VTI) in the range 1.45-300 K and helium gas atmosphere. Resistivity was extracted from four-probe measurements with current values of order of 10 $\mu$A to avoid any Joule heating.


\bibliography{bibliography}

\begin{thebibliography}{45}%
\makeatletter
\providecommand \@ifxundefined [1]{%
 \@ifx{#1\undefined}
}%
\providecommand \@ifnum [1]{%
 \ifnum #1\expandafter \@firstoftwo
 \else \expandafter \@secondoftwo
 \fi
}%
\providecommand \@ifx [1]{%
 \ifx #1\expandafter \@firstoftwo
 \else \expandafter \@secondoftwo
 \fi
}%
\providecommand \natexlab [1]{#1}%
\providecommand \enquote  [1]{``#1''}%
\providecommand \bibnamefont  [1]{#1}%
\providecommand \bibfnamefont [1]{#1}%
\providecommand \citenamefont [1]{#1}%
\providecommand \href@noop [0]{\@secondoftwo}%
\providecommand \href [0]{\begingroup \@sanitize@url \@href}%
\providecommand \@href[1]{\@@startlink{#1}\@@href}%
\providecommand \@@href[1]{\endgroup#1\@@endlink}%
\providecommand \@sanitize@url [0]{\catcode `\\12\catcode `\$12\catcode
  `\&12\catcode `\#12\catcode `\^12\catcode `\_12\catcode `\%12\relax}%
\providecommand \@@startlink[1]{}%
\providecommand \@@endlink[0]{}%
\providecommand \url  [0]{\begingroup\@sanitize@url \@url }%
\providecommand \@url [1]{\endgroup\@href {#1}{\urlprefix }}%
\providecommand \urlprefix  [0]{URL }%
\providecommand \Eprint [0]{\href }%
\providecommand \doibase [0]{https://doi.org/}%
\providecommand \selectlanguage [0]{\@gobble}%
\providecommand \bibinfo  [0]{\@secondoftwo}%
\providecommand \bibfield  [0]{\@secondoftwo}%
\providecommand \translation [1]{[#1]}%
\providecommand \BibitemOpen [0]{}%
\providecommand \bibitemStop [0]{}%
\providecommand \bibitemNoStop [0]{.\EOS\space}%
\providecommand \EOS [0]{\spacefactor3000\relax}%
\providecommand \BibitemShut  [1]{\csname bibitem#1\endcsname}%
\let\auto@bib@innerbib\@empty
\bibitem [{\citenamefont {Novoselov}\ \emph {et~al.}(2007)\citenamefont
  {Novoselov}, \citenamefont {Jiang}, \citenamefont {Zhang}, \citenamefont
  {Morozov}, \citenamefont {Stormer}, \citenamefont {Zeitler}, \citenamefont
  {Maan}, \citenamefont {Boebinger}, \citenamefont {Kim},\ and\ \citenamefont
  {Geim}}]{Graphene}%
  \BibitemOpen
  \bibfield  {author} {\bibinfo {author} {\bibfnamefont {K.~S.}\ \bibnamefont
  {Novoselov}}, \bibinfo {author} {\bibfnamefont {Z.}~\bibnamefont {Jiang}},
  \bibinfo {author} {\bibfnamefont {Y.}~\bibnamefont {Zhang}}, \bibinfo
  {author} {\bibfnamefont {S.~V.}\ \bibnamefont {Morozov}}, \bibinfo {author}
  {\bibfnamefont {H.~L.}\ \bibnamefont {Stormer}}, \bibinfo {author}
  {\bibfnamefont {U.}~\bibnamefont {Zeitler}}, \bibinfo {author} {\bibfnamefont
  {J.~C.}\ \bibnamefont {Maan}}, \bibinfo {author} {\bibfnamefont {G.~S.}\
  \bibnamefont {Boebinger}}, \bibinfo {author} {\bibfnamefont {P.}~\bibnamefont
  {Kim}},\ and\ \bibinfo {author} {\bibfnamefont {A.~K.}\ \bibnamefont
  {Geim}},\ }\bibfield  {title} {\bibinfo {title} {Room-temperature quantum
  {Hall} effect in graphene},\ }\href {https://doi.org/10.1126/science.1137201}
  {\bibfield  {journal} {\bibinfo  {journal} {Science}\ }\textbf {\bibinfo
  {volume} {315}},\ \bibinfo {pages} {1379} (\bibinfo {year}
  {2007})}\BibitemShut {NoStop}%
\bibitem [{\citenamefont {Ci}\ \emph {et~al.}(2010)\citenamefont {Ci},
  \citenamefont {Song}, \citenamefont {Jin}, \citenamefont {Jariwala},
  \citenamefont {Wu}, \citenamefont {Li}, \citenamefont {Srivastava},
  \citenamefont {Wang}, \citenamefont {Storr}, \citenamefont {Balicas},
  \citenamefont {Liu},\ and\ \citenamefont {Ajayan}}]{Layered_BP}%
  \BibitemOpen
  \bibfield  {author} {\bibinfo {author} {\bibfnamefont {L.}~\bibnamefont
  {Ci}}, \bibinfo {author} {\bibfnamefont {L.}~\bibnamefont {Song}}, \bibinfo
  {author} {\bibfnamefont {C.}~\bibnamefont {Jin}}, \bibinfo {author}
  {\bibfnamefont {D.}~\bibnamefont {Jariwala}}, \bibinfo {author}
  {\bibfnamefont {D.}~\bibnamefont {Wu}}, \bibinfo {author} {\bibfnamefont
  {Y.}~\bibnamefont {Li}}, \bibinfo {author} {\bibfnamefont {A.}~\bibnamefont
  {Srivastava}}, \bibinfo {author} {\bibfnamefont {Z.~F.}\ \bibnamefont
  {Wang}}, \bibinfo {author} {\bibfnamefont {K.}~\bibnamefont {Storr}},
  \bibinfo {author} {\bibfnamefont {L.}~\bibnamefont {Balicas}}, \bibinfo
  {author} {\bibfnamefont {F.}~\bibnamefont {Liu}},\ and\ \bibinfo {author}
  {\bibfnamefont {P.~M.}\ \bibnamefont {Ajayan}},\ }\bibfield  {title}
  {\bibinfo {title} {Atomic layers of hybridized boron nitride and graphene
  domains},\ }\href {https://doi.org/10.1038/nmat2711} {\bibfield  {journal}
  {\bibinfo  {journal} {Nature Materials}\ }\textbf {\bibinfo {volume} {9}},\
  \bibinfo {pages} {430} (\bibinfo {year} {2010})}\BibitemShut {NoStop}%
\bibitem [{\citenamefont {Ai}\ \emph {et~al.}(2017)\citenamefont {Ai},
  \citenamefont {Guan}, \citenamefont {Li}, \citenamefont {Yao}, \citenamefont
  {Chen}, \citenamefont {Zhang}, \citenamefont {Duan},\ and\ \citenamefont
  {Duan}}]{Layered_Halides_1}%
  \BibitemOpen
  \bibfield  {author} {\bibinfo {author} {\bibfnamefont {R.}~\bibnamefont
  {Ai}}, \bibinfo {author} {\bibfnamefont {X.}~\bibnamefont {Guan}}, \bibinfo
  {author} {\bibfnamefont {J.}~\bibnamefont {Li}}, \bibinfo {author}
  {\bibfnamefont {K.}~\bibnamefont {Yao}}, \bibinfo {author} {\bibfnamefont
  {P.}~\bibnamefont {Chen}}, \bibinfo {author} {\bibfnamefont {Z.}~\bibnamefont
  {Zhang}}, \bibinfo {author} {\bibfnamefont {X.}~\bibnamefont {Duan}},\ and\
  \bibinfo {author} {\bibfnamefont {X.}~\bibnamefont {Duan}},\ }\bibfield
  {title} {\bibinfo {title} {Growth of single-crystalline cadmium iodide
  nanoplates, {CdI$_{2}$/MoS$_{2}$ (WS$_{2}$, WSe$_{2}$)} van der waals
  heterostructures, and patterned arrays},\ }\href
  {https://doi.org/10.1021/acsnano.7b01507} {\bibfield  {journal} {\bibinfo
  {journal} {ACS Nano}\ }\textbf {\bibinfo {volume} {11}},\ \bibinfo {pages}
  {3413} (\bibinfo {year} {2017})}\BibitemShut {NoStop}%
\bibitem [{\citenamefont {Li}\ \emph {et~al.}(2017)\citenamefont {Li},
  \citenamefont {Guan}, \citenamefont {Wang}, \citenamefont {Cheng},
  \citenamefont {Ai}, \citenamefont {Yao}, \citenamefont {Chen}, \citenamefont
  {Zhang}, \citenamefont {Duan},\ and\ \citenamefont
  {Duan}}]{Layered_Halides_2}%
  \BibitemOpen
  \bibfield  {author} {\bibinfo {author} {\bibfnamefont {J.}~\bibnamefont
  {Li}}, \bibinfo {author} {\bibfnamefont {X.}~\bibnamefont {Guan}}, \bibinfo
  {author} {\bibfnamefont {C.}~\bibnamefont {Wang}}, \bibinfo {author}
  {\bibfnamefont {H.-C.}\ \bibnamefont {Cheng}}, \bibinfo {author}
  {\bibfnamefont {R.}~\bibnamefont {Ai}}, \bibinfo {author} {\bibfnamefont
  {K.}~\bibnamefont {Yao}}, \bibinfo {author} {\bibfnamefont {P.}~\bibnamefont
  {Chen}}, \bibinfo {author} {\bibfnamefont {Z.}~\bibnamefont {Zhang}},
  \bibinfo {author} {\bibfnamefont {X.}~\bibnamefont {Duan}},\ and\ \bibinfo
  {author} {\bibfnamefont {X.}~\bibnamefont {Duan}},\ }\bibfield  {title}
  {\bibinfo {title} {Synthesis of {2D} layered {BiI$_{3}$} nanoplates,
  {BiI$_{3}$/WSe$_{2}$} van der waals heterostructures and their electronic,
  optoelectronic properties},\ }\href
  {https://doi.org/https://doi.org/10.1002/smll.201701034} {\bibfield
  {journal} {\bibinfo  {journal} {Small}\ }\textbf {\bibinfo {volume} {13}},\
  \bibinfo {pages} {1701034} (\bibinfo {year} {2017})}\BibitemShut {NoStop}%
\bibitem [{\citenamefont {Kolobov}\ and\ \citenamefont
  {Tominaga}(2017)}]{2D_TMDC_book}%
  \BibitemOpen
  \bibfield  {author} {\bibinfo {author} {\bibfnamefont {A.~V.}\ \bibnamefont
  {Kolobov}}\ and\ \bibinfo {author} {\bibfnamefont {J.}~\bibnamefont
  {Tominaga}},\ }\bibfield  {title} {\bibinfo {title} {Two-dimensional
  transition-metal dichalcogenides},\ }\href
  {https://doi.org/10.1557/mrs.2017.135} {\bibfield  {journal} {\bibinfo
  {journal} {MRS Bulletin}\ }\textbf {\bibinfo {volume} {42}},\ \bibinfo
  {pages} {471} (\bibinfo {year} {2017})}\BibitemShut {NoStop}%
\bibitem [{\citenamefont {Chhowalla}\ \emph {et~al.}(2013)\citenamefont
  {Chhowalla}, \citenamefont {Shin}, \citenamefont {Eda}, \citenamefont {Li},
  \citenamefont {Loh},\ and\ \citenamefont {Zhang}}]{TMDC_general}%
  \BibitemOpen
  \bibfield  {author} {\bibinfo {author} {\bibfnamefont {M.}~\bibnamefont
  {Chhowalla}}, \bibinfo {author} {\bibfnamefont {H.~S.}\ \bibnamefont {Shin}},
  \bibinfo {author} {\bibfnamefont {G.}~\bibnamefont {Eda}}, \bibinfo {author}
  {\bibfnamefont {L.-J.}\ \bibnamefont {Li}}, \bibinfo {author} {\bibfnamefont
  {K.~P.}\ \bibnamefont {Loh}},\ and\ \bibinfo {author} {\bibfnamefont
  {H.}~\bibnamefont {Zhang}},\ }\bibfield  {title} {\bibinfo {title} {The
  chemistry of two-dimensional layered transition metal dichalcogenide
  nanosheets},\ }\href {https://doi.org/10.1038/nchem.1589} {\bibfield
  {journal} {\bibinfo  {journal} {Nature Chemistry}\ }\textbf {\bibinfo
  {volume} {5}},\ \bibinfo {pages} {263} (\bibinfo {year} {2013})}\BibitemShut
  {NoStop}%
\bibitem [{\citenamefont {Ye}\ \emph {et~al.}(2015)\citenamefont {Ye},
  \citenamefont {Wong}, \citenamefont {Lu}, \citenamefont {Ni}, \citenamefont
  {Zhu}, \citenamefont {Chen}, \citenamefont {Wang},\ and\ \citenamefont
  {Zhang}}]{TMDC_laser_1}%
  \BibitemOpen
  \bibfield  {author} {\bibinfo {author} {\bibfnamefont {Y.}~\bibnamefont
  {Ye}}, \bibinfo {author} {\bibfnamefont {Z.~J.}\ \bibnamefont {Wong}},
  \bibinfo {author} {\bibfnamefont {X.}~\bibnamefont {Lu}}, \bibinfo {author}
  {\bibfnamefont {X.}~\bibnamefont {Ni}}, \bibinfo {author} {\bibfnamefont
  {H.}~\bibnamefont {Zhu}}, \bibinfo {author} {\bibfnamefont {X.}~\bibnamefont
  {Chen}}, \bibinfo {author} {\bibfnamefont {Y.}~\bibnamefont {Wang}},\ and\
  \bibinfo {author} {\bibfnamefont {X.}~\bibnamefont {Zhang}},\ }\bibfield
  {title} {\bibinfo {title} {Monolayer excitonic laser},\ }\href
  {https://doi.org/10.1038/nphoton.2015.197} {\bibfield  {journal} {\bibinfo
  {journal} {Nature Photonics}\ }\textbf {\bibinfo {volume} {9}},\ \bibinfo
  {pages} {733} (\bibinfo {year} {2015})}\BibitemShut {NoStop}%
\bibitem [{\citenamefont {Ge}\ \emph {et~al.}(2019)\citenamefont {Ge},
  \citenamefont {Minkov}, \citenamefont {Fan}, \citenamefont {Li},\ and\
  \citenamefont {Zhou}}]{TMDC_laser_2}%
  \BibitemOpen
  \bibfield  {author} {\bibinfo {author} {\bibfnamefont {X.}~\bibnamefont
  {Ge}}, \bibinfo {author} {\bibfnamefont {M.}~\bibnamefont {Minkov}}, \bibinfo
  {author} {\bibfnamefont {S.}~\bibnamefont {Fan}}, \bibinfo {author}
  {\bibfnamefont {X.}~\bibnamefont {Li}},\ and\ \bibinfo {author}
  {\bibfnamefont {W.}~\bibnamefont {Zhou}},\ }\bibfield  {title} {\bibinfo
  {title} {Laterally confined photonic crystal surface emitting laser
  incorporating monolayer tungsten disulfide},\ }\href
  {https://doi.org/10.1038/s41699-019-0099-1} {\bibfield  {journal} {\bibinfo
  {journal} {npj {2D} Materials and Applications}\ }\textbf {\bibinfo {volume}
  {3}},\ \bibinfo {pages} {16} (\bibinfo {year} {2019})}\BibitemShut {NoStop}%
\bibitem [{\citenamefont {Chee}\ \emph {et~al.}(2020)\citenamefont {Chee},
  \citenamefont {Lee}, \citenamefont {Jo}, \citenamefont {Cho}, \citenamefont
  {Chun}, \citenamefont {Baik}, \citenamefont {Kim}, \citenamefont {Yoon},
  \citenamefont {Lee},\ and\ \citenamefont {Ham}}]{Nanoelectronics_1}%
  \BibitemOpen
  \bibfield  {author} {\bibinfo {author} {\bibfnamefont {S.-S.}\ \bibnamefont
  {Chee}}, \bibinfo {author} {\bibfnamefont {W.-J.}\ \bibnamefont {Lee}},
  \bibinfo {author} {\bibfnamefont {Y.-R.}\ \bibnamefont {Jo}}, \bibinfo
  {author} {\bibfnamefont {M.~K.}\ \bibnamefont {Cho}}, \bibinfo {author}
  {\bibfnamefont {D.}~\bibnamefont {Chun}}, \bibinfo {author} {\bibfnamefont
  {H.}~\bibnamefont {Baik}}, \bibinfo {author} {\bibfnamefont {B.-J.}\
  \bibnamefont {Kim}}, \bibinfo {author} {\bibfnamefont {M.-H.}\ \bibnamefont
  {Yoon}}, \bibinfo {author} {\bibfnamefont {K.}~\bibnamefont {Lee}},\ and\
  \bibinfo {author} {\bibfnamefont {M.-H.}\ \bibnamefont {Ham}},\ }\bibfield
  {title} {\bibinfo {title} {Atomic vacancy control and elemental substitution
  in a monolayer molybdenum disulfide for high performance optoelectronic
  device arrays},\ }\href
  {https://doi.org/https://doi.org/10.1002/adfm.201908147} {\bibfield
  {journal} {\bibinfo  {journal} {Advanced Functional Materials}\ }\textbf
  {\bibinfo {volume} {30}},\ \bibinfo {pages} {1908147} (\bibinfo {year}
  {2020})}\BibitemShut {NoStop}%
\bibitem [{\citenamefont {Wang}\ \emph {et~al.}(2020)\citenamefont {Wang},
  \citenamefont {Wang}, \citenamefont {Wang}, \citenamefont {Ye}, \citenamefont
  {He}, \citenamefont {Wu}, \citenamefont {Peng}, \citenamefont {Wu},
  \citenamefont {Chen}, \citenamefont {Zhong}, \citenamefont {Xie},
  \citenamefont {Cui}, \citenamefont {Shen}, \citenamefont {Zhang},
  \citenamefont {Gu}, \citenamefont {Luo}, \citenamefont {Wang}, \citenamefont
  {Chen}, \citenamefont {Zhou}, \citenamefont {Pan}, \citenamefont {Zhou},
  \citenamefont {Zhang},\ and\ \citenamefont {Hu}}]{Nanoelectronics_2}%
  \BibitemOpen
  \bibfield  {author} {\bibinfo {author} {\bibfnamefont {Z.}~\bibnamefont
  {Wang}}, \bibinfo {author} {\bibfnamefont {P.}~\bibnamefont {Wang}}, \bibinfo
  {author} {\bibfnamefont {F.}~\bibnamefont {Wang}}, \bibinfo {author}
  {\bibfnamefont {J.}~\bibnamefont {Ye}}, \bibinfo {author} {\bibfnamefont
  {T.}~\bibnamefont {He}}, \bibinfo {author} {\bibfnamefont {F.}~\bibnamefont
  {Wu}}, \bibinfo {author} {\bibfnamefont {M.}~\bibnamefont {Peng}}, \bibinfo
  {author} {\bibfnamefont {P.}~\bibnamefont {Wu}}, \bibinfo {author}
  {\bibfnamefont {Y.}~\bibnamefont {Chen}}, \bibinfo {author} {\bibfnamefont
  {F.}~\bibnamefont {Zhong}}, \bibinfo {author} {\bibfnamefont
  {R.}~\bibnamefont {Xie}}, \bibinfo {author} {\bibfnamefont {Z.}~\bibnamefont
  {Cui}}, \bibinfo {author} {\bibfnamefont {L.}~\bibnamefont {Shen}}, \bibinfo
  {author} {\bibfnamefont {Q.}~\bibnamefont {Zhang}}, \bibinfo {author}
  {\bibfnamefont {L.}~\bibnamefont {Gu}}, \bibinfo {author} {\bibfnamefont
  {M.}~\bibnamefont {Luo}}, \bibinfo {author} {\bibfnamefont {Y.}~\bibnamefont
  {Wang}}, \bibinfo {author} {\bibfnamefont {H.}~\bibnamefont {Chen}}, \bibinfo
  {author} {\bibfnamefont {P.}~\bibnamefont {Zhou}}, \bibinfo {author}
  {\bibfnamefont {A.}~\bibnamefont {Pan}}, \bibinfo {author} {\bibfnamefont
  {X.}~\bibnamefont {Zhou}}, \bibinfo {author} {\bibfnamefont {L.}~\bibnamefont
  {Zhang}},\ and\ \bibinfo {author} {\bibfnamefont {W.}~\bibnamefont {Hu}},\
  }\bibfield  {title} {\bibinfo {title} {A noble metal dichalcogenide for
  high-performance field-effect transistors and broadband photodetectors},\
  }\href {https://doi.org/https://doi.org/10.1002/adfm.201907945} {\bibfield
  {journal} {\bibinfo  {journal} {Advanced Functional Materials}\ }\textbf
  {\bibinfo {volume} {30}},\ \bibinfo {pages} {1907945} (\bibinfo {year}
  {2020})}\BibitemShut {NoStop}%
\bibitem [{\citenamefont {Cui}\ \emph {et~al.}(2020)\citenamefont {Cui},
  \citenamefont {Liu}, \citenamefont {Feng}, \citenamefont {Zhang},
  \citenamefont {Du}, \citenamefont {Liu}, \citenamefont {Wang}, \citenamefont
  {Chen},\ and\ \citenamefont {Zhou}}]{EnergyStorage_1}%
  \BibitemOpen
  \bibfield  {author} {\bibinfo {author} {\bibfnamefont {Y.}~\bibnamefont
  {Cui}}, \bibinfo {author} {\bibfnamefont {W.}~\bibnamefont {Liu}}, \bibinfo
  {author} {\bibfnamefont {W.}~\bibnamefont {Feng}}, \bibinfo {author}
  {\bibfnamefont {Y.}~\bibnamefont {Zhang}}, \bibinfo {author} {\bibfnamefont
  {Y.}~\bibnamefont {Du}}, \bibinfo {author} {\bibfnamefont {S.}~\bibnamefont
  {Liu}}, \bibinfo {author} {\bibfnamefont {H.}~\bibnamefont {Wang}}, \bibinfo
  {author} {\bibfnamefont {M.}~\bibnamefont {Chen}},\ and\ \bibinfo {author}
  {\bibfnamefont {J.}~\bibnamefont {Zhou}},\ }\bibfield  {title} {\bibinfo
  {title} {Controlled design of well-dispersed ultrathin {MoS$_{2}$} nanosheets
  inside hollow carbon skeleton: Toward fast potassium storage by constructing
  spacious “houses” for k ions},\ }\href
  {https://doi.org/https://doi.org/10.1002/adfm.201908755} {\bibfield
  {journal} {\bibinfo  {journal} {Advanced Functional Materials}\ }\textbf
  {\bibinfo {volume} {30}},\ \bibinfo {pages} {1908755} (\bibinfo {year}
  {2020})}\BibitemShut {NoStop}%
\bibitem [{\citenamefont {Shi}\ \emph {et~al.}(2020{\natexlab{a}})\citenamefont
  {Shi}, \citenamefont {Sun}, \citenamefont {Yin}, \citenamefont {Shen},
  \citenamefont {Shi}, \citenamefont {Zhao}, \citenamefont {Zhao},\ and\
  \citenamefont {Zhang}}]{EnergyStorage_2}%
  \BibitemOpen
  \bibfield  {author} {\bibinfo {author} {\bibfnamefont {S.}~\bibnamefont
  {Shi}}, \bibinfo {author} {\bibfnamefont {C.}~\bibnamefont {Sun}}, \bibinfo
  {author} {\bibfnamefont {X.}~\bibnamefont {Yin}}, \bibinfo {author}
  {\bibfnamefont {L.}~\bibnamefont {Shen}}, \bibinfo {author} {\bibfnamefont
  {Q.}~\bibnamefont {Shi}}, \bibinfo {author} {\bibfnamefont {K.}~\bibnamefont
  {Zhao}}, \bibinfo {author} {\bibfnamefont {Y.}~\bibnamefont {Zhao}},\ and\
  \bibinfo {author} {\bibfnamefont {J.}~\bibnamefont {Zhang}},\ }\bibfield
  {title} {\bibinfo {title} {Fep quantum dots confined in
  carbon-nanotube-grafted p-doped carbon octahedra for high-rate sodium storage
  and full-cell applications},\ }\href
  {https://doi.org/https://doi.org/10.1002/adfm.201909283} {\bibfield
  {journal} {\bibinfo  {journal} {Advanced Functional Materials}\ }\textbf
  {\bibinfo {volume} {30}},\ \bibinfo {pages} {1909283} (\bibinfo {year}
  {2020}{\natexlab{a}})}\BibitemShut {NoStop}%
\bibitem [{\citenamefont {Tay}\ \emph {et~al.}(2020)\citenamefont {Tay},
  \citenamefont {Li}, \citenamefont {Lin}, \citenamefont {Wang}, \citenamefont
  {Lim}, \citenamefont {Chen}, \citenamefont {Leong}, \citenamefont {Tsang},\
  and\ \citenamefont {Teo}}]{Sensing}%
  \BibitemOpen
  \bibfield  {author} {\bibinfo {author} {\bibfnamefont {R.~Y.}\ \bibnamefont
  {Tay}}, \bibinfo {author} {\bibfnamefont {H.}~\bibnamefont {Li}}, \bibinfo
  {author} {\bibfnamefont {J.}~\bibnamefont {Lin}}, \bibinfo {author}
  {\bibfnamefont {H.}~\bibnamefont {Wang}}, \bibinfo {author} {\bibfnamefont
  {J.~S.~K.}\ \bibnamefont {Lim}}, \bibinfo {author} {\bibfnamefont
  {S.}~\bibnamefont {Chen}}, \bibinfo {author} {\bibfnamefont {W.~L.}\
  \bibnamefont {Leong}}, \bibinfo {author} {\bibfnamefont {S.~H.}\ \bibnamefont
  {Tsang}},\ and\ \bibinfo {author} {\bibfnamefont {E.~H.~T.}\ \bibnamefont
  {Teo}},\ }\bibfield  {title} {\bibinfo {title} {Lightweight, superelastic
  boron nitride/polydimethylsiloxane foam as air dielectric substitute for
  multifunctional capacitive sensor applications},\ }\href
  {https://doi.org/https://doi.org/10.1002/adfm.201909604} {\bibfield
  {journal} {\bibinfo  {journal} {Advanced Functional Materials}\ }\textbf
  {\bibinfo {volume} {30}},\ \bibinfo {pages} {1909604} (\bibinfo {year}
  {2020})}\BibitemShut {NoStop}%
\bibitem [{\citenamefont {Yan}\ \emph {et~al.}(2017)\citenamefont {Yan},
  \citenamefont {Huang}, \citenamefont {Zhang}, \citenamefont {Wang},
  \citenamefont {Yao}, \citenamefont {Deng}, \citenamefont {Wan}, \citenamefont
  {Zhang}, \citenamefont {Arita}, \citenamefont {Yang}, \citenamefont {Sun},
  \citenamefont {Yao}, \citenamefont {Wu}, \citenamefont {Fan}, \citenamefont
  {Duan},\ and\ \citenamefont {Zhou}}]{PtTe2_Dirac}%
  \BibitemOpen
  \bibfield  {author} {\bibinfo {author} {\bibfnamefont {M.}~\bibnamefont
  {Yan}}, \bibinfo {author} {\bibfnamefont {H.}~\bibnamefont {Huang}}, \bibinfo
  {author} {\bibfnamefont {K.}~\bibnamefont {Zhang}}, \bibinfo {author}
  {\bibfnamefont {E.}~\bibnamefont {Wang}}, \bibinfo {author} {\bibfnamefont
  {W.}~\bibnamefont {Yao}}, \bibinfo {author} {\bibfnamefont {K.}~\bibnamefont
  {Deng}}, \bibinfo {author} {\bibfnamefont {G.}~\bibnamefont {Wan}}, \bibinfo
  {author} {\bibfnamefont {H.}~\bibnamefont {Zhang}}, \bibinfo {author}
  {\bibfnamefont {M.}~\bibnamefont {Arita}}, \bibinfo {author} {\bibfnamefont
  {H.}~\bibnamefont {Yang}}, \bibinfo {author} {\bibfnamefont {Z.}~\bibnamefont
  {Sun}}, \bibinfo {author} {\bibfnamefont {H.}~\bibnamefont {Yao}}, \bibinfo
  {author} {\bibfnamefont {Y.}~\bibnamefont {Wu}}, \bibinfo {author}
  {\bibfnamefont {S.}~\bibnamefont {Fan}}, \bibinfo {author} {\bibfnamefont
  {W.}~\bibnamefont {Duan}},\ and\ \bibinfo {author} {\bibfnamefont
  {S.}~\bibnamefont {Zhou}},\ }\bibfield  {title} {\bibinfo {title}
  {Lorentz-violating type-{II} {Dirac} fermions in transition metal
  dichalcogenide {PtTe$_{2}$}},\ }\href
  {https://doi.org/10.1038/s41467-017-00280-6} {\bibfield  {journal} {\bibinfo
  {journal} {Nature Communications}\ }\textbf {\bibinfo {volume} {8}},\
  \bibinfo {pages} {257} (\bibinfo {year} {2017})}\BibitemShut {NoStop}%
\bibitem [{\citenamefont {Jiang}\ \emph {et~al.}(2017)\citenamefont {Jiang},
  \citenamefont {Liu}, \citenamefont {Sun}, \citenamefont {Yang}, \citenamefont
  {Rajamathi}, \citenamefont {Qi}, \citenamefont {Yang}, \citenamefont {Chen},
  \citenamefont {Peng}, \citenamefont {Hwang}, \citenamefont {Sun},
  \citenamefont {Mo}, \citenamefont {Vobornik}, \citenamefont {Fujii},
  \citenamefont {Parkin}, \citenamefont {Felser}, \citenamefont {Yan},\ and\
  \citenamefont {Chen}}]{MoTe2_Weyl}%
  \BibitemOpen
  \bibfield  {author} {\bibinfo {author} {\bibfnamefont {J.}~\bibnamefont
  {Jiang}}, \bibinfo {author} {\bibfnamefont {Z.~K.}\ \bibnamefont {Liu}},
  \bibinfo {author} {\bibfnamefont {Y.}~\bibnamefont {Sun}}, \bibinfo {author}
  {\bibfnamefont {H.~F.}\ \bibnamefont {Yang}}, \bibinfo {author}
  {\bibfnamefont {C.~R.}\ \bibnamefont {Rajamathi}}, \bibinfo {author}
  {\bibfnamefont {Y.~P.}\ \bibnamefont {Qi}}, \bibinfo {author} {\bibfnamefont
  {L.~X.}\ \bibnamefont {Yang}}, \bibinfo {author} {\bibfnamefont
  {C.}~\bibnamefont {Chen}}, \bibinfo {author} {\bibfnamefont {H.}~\bibnamefont
  {Peng}}, \bibinfo {author} {\bibfnamefont {C.-C.}\ \bibnamefont {Hwang}},
  \bibinfo {author} {\bibfnamefont {S.~Z.}\ \bibnamefont {Sun}}, \bibinfo
  {author} {\bibfnamefont {S.-K.}\ \bibnamefont {Mo}}, \bibinfo {author}
  {\bibfnamefont {I.}~\bibnamefont {Vobornik}}, \bibinfo {author}
  {\bibfnamefont {J.}~\bibnamefont {Fujii}}, \bibinfo {author} {\bibfnamefont
  {S.~S.~P.}\ \bibnamefont {Parkin}}, \bibinfo {author} {\bibfnamefont
  {C.}~\bibnamefont {Felser}}, \bibinfo {author} {\bibfnamefont {B.~H.}\
  \bibnamefont {Yan}},\ and\ \bibinfo {author} {\bibfnamefont {Y.~L.}\
  \bibnamefont {Chen}},\ }\bibfield  {title} {\bibinfo {title} {Signature of
  type-{II} weyl semimetal phase in {MoTe$_{2}$}},\ }\href
  {https://doi.org/10.1038/ncomms13973} {\bibfield  {journal} {\bibinfo
  {journal} {Nature Communications}\ }\textbf {\bibinfo {volume} {8}},\
  \bibinfo {pages} {13973} (\bibinfo {year} {2017})}\BibitemShut {NoStop}%
\bibitem [{\citenamefont {Xu}\ \emph {et~al.}(2018)\citenamefont {Xu},
  \citenamefont {Li}, \citenamefont {Jiao}, \citenamefont {Zhou}, \citenamefont
  {Qian}, \citenamefont {Sankar}, \citenamefont {Zhigadlo}, \citenamefont {Qi},
  \citenamefont {Qian}, \citenamefont {Chou},\ and\ \citenamefont
  {Xu}}]{ACS_type2}%
  \BibitemOpen
  \bibfield  {author} {\bibinfo {author} {\bibfnamefont {C.}~\bibnamefont
  {Xu}}, \bibinfo {author} {\bibfnamefont {B.}~\bibnamefont {Li}}, \bibinfo
  {author} {\bibfnamefont {W.}~\bibnamefont {Jiao}}, \bibinfo {author}
  {\bibfnamefont {W.}~\bibnamefont {Zhou}}, \bibinfo {author} {\bibfnamefont
  {B.}~\bibnamefont {Qian}}, \bibinfo {author} {\bibfnamefont {R.}~\bibnamefont
  {Sankar}}, \bibinfo {author} {\bibfnamefont {N.~D.}\ \bibnamefont
  {Zhigadlo}}, \bibinfo {author} {\bibfnamefont {Y.}~\bibnamefont {Qi}},
  \bibinfo {author} {\bibfnamefont {D.}~\bibnamefont {Qian}}, \bibinfo {author}
  {\bibfnamefont {F.-C.}\ \bibnamefont {Chou}},\ and\ \bibinfo {author}
  {\bibfnamefont {X.}~\bibnamefont {Xu}},\ }\bibfield  {title} {\bibinfo
  {title} {Topological type-{II} dirac fermions approaching the fermi level in
  a transition metal dichalcogenide {NiTe$_{2}$}},\ }\href
  {https://doi.org/10.1021/acs.chemmater.8b02132} {\bibfield  {journal}
  {\bibinfo  {journal} {Chemistry of Materials}\ }\textbf {\bibinfo {volume}
  {30}},\ \bibinfo {pages} {4823} (\bibinfo {year} {2018})}\BibitemShut
  {NoStop}%
\bibitem [{\citenamefont {Nappini}\ \emph {et~al.}(2020)\citenamefont
  {Nappini}, \citenamefont {Boukhvalov}, \citenamefont {D'Olimpio},
  \citenamefont {Zhang}, \citenamefont {Ghosh}, \citenamefont {Kuo},
  \citenamefont {Zhu}, \citenamefont {Cheng}, \citenamefont {Nardone},
  \citenamefont {Ottaviano}, \citenamefont {Mondal}, \citenamefont {Edla},
  \citenamefont {Fuji}, \citenamefont {Lue}, \citenamefont {Vobornik},
  \citenamefont {Yarmoff}, \citenamefont {Agarwal}, \citenamefont {Wang},
  \citenamefont {Zhang}, \citenamefont {Bondino},\ and\ \citenamefont
  {Politano}}]{exfoliated}%
  \BibitemOpen
  \bibfield  {author} {\bibinfo {author} {\bibfnamefont {S.}~\bibnamefont
  {Nappini}}, \bibinfo {author} {\bibfnamefont {D.~W.}\ \bibnamefont
  {Boukhvalov}}, \bibinfo {author} {\bibfnamefont {G.}~\bibnamefont
  {D'Olimpio}}, \bibinfo {author} {\bibfnamefont {L.}~\bibnamefont {Zhang}},
  \bibinfo {author} {\bibfnamefont {B.}~\bibnamefont {Ghosh}}, \bibinfo
  {author} {\bibfnamefont {C.-N.}\ \bibnamefont {Kuo}}, \bibinfo {author}
  {\bibfnamefont {H.}~\bibnamefont {Zhu}}, \bibinfo {author} {\bibfnamefont
  {J.}~\bibnamefont {Cheng}}, \bibinfo {author} {\bibfnamefont
  {M.}~\bibnamefont {Nardone}}, \bibinfo {author} {\bibfnamefont
  {L.}~\bibnamefont {Ottaviano}}, \bibinfo {author} {\bibfnamefont
  {D.}~\bibnamefont {Mondal}}, \bibinfo {author} {\bibfnamefont
  {R.}~\bibnamefont {Edla}}, \bibinfo {author} {\bibfnamefont {J.}~\bibnamefont
  {Fuji}}, \bibinfo {author} {\bibfnamefont {C.~S.}\ \bibnamefont {Lue}},
  \bibinfo {author} {\bibfnamefont {I.}~\bibnamefont {Vobornik}}, \bibinfo
  {author} {\bibfnamefont {J.~A.}\ \bibnamefont {Yarmoff}}, \bibinfo {author}
  {\bibfnamefont {A.}~\bibnamefont {Agarwal}}, \bibinfo {author} {\bibfnamefont
  {L.}~\bibnamefont {Wang}}, \bibinfo {author} {\bibfnamefont {L.}~\bibnamefont
  {Zhang}}, \bibinfo {author} {\bibfnamefont {F.}~\bibnamefont {Bondino}},\
  and\ \bibinfo {author} {\bibfnamefont {A.}~\bibnamefont {Politano}},\
  }\bibfield  {title} {\bibinfo {title} {Transition-metal dichalcogenide
  {NiTe$_{2}$}: An ambient-stable material for catalysis and nanoelectronics},\
  }\href {https://doi.org/10.1002/adfm.202000915} {\bibfield  {journal}
  {\bibinfo  {journal} {Advanced Functional Materials}\ }\textbf {\bibinfo
  {volume} {30}},\ \bibinfo {pages} {2000915} (\bibinfo {year}
  {2020})}\BibitemShut {NoStop}%
\bibitem [{\citenamefont {Zhang}\ \emph {et~al.}(2021)\citenamefont {Zhang},
  \citenamefont {Chen}, \citenamefont {Zhang}, \citenamefont {Wang},
  \citenamefont {Xu}, \citenamefont {Han}, \citenamefont {Guo}, \citenamefont
  {Yang}, \citenamefont {Kuo}, \citenamefont {Lue}, \citenamefont {Mondal},
  \citenamefont {Fuji}, \citenamefont {Vobornik}, \citenamefont {Ghosh},
  \citenamefont {Agarwal}, \citenamefont {Xing}, \citenamefont {Chen},
  \citenamefont {Politano},\ and\ \citenamefont {Lu}}]{NiTe2_THz_1}%
  \BibitemOpen
  \bibfield  {author} {\bibinfo {author} {\bibfnamefont {L.}~\bibnamefont
  {Zhang}}, \bibinfo {author} {\bibfnamefont {Z.}~\bibnamefont {Chen}},
  \bibinfo {author} {\bibfnamefont {K.}~\bibnamefont {Zhang}}, \bibinfo
  {author} {\bibfnamefont {L.}~\bibnamefont {Wang}}, \bibinfo {author}
  {\bibfnamefont {H.}~\bibnamefont {Xu}}, \bibinfo {author} {\bibfnamefont
  {L.}~\bibnamefont {Han}}, \bibinfo {author} {\bibfnamefont {W.}~\bibnamefont
  {Guo}}, \bibinfo {author} {\bibfnamefont {Y.}~\bibnamefont {Yang}}, \bibinfo
  {author} {\bibfnamefont {C.-N.}\ \bibnamefont {Kuo}}, \bibinfo {author}
  {\bibfnamefont {C.~S.}\ \bibnamefont {Lue}}, \bibinfo {author} {\bibfnamefont
  {D.}~\bibnamefont {Mondal}}, \bibinfo {author} {\bibfnamefont
  {J.}~\bibnamefont {Fuji}}, \bibinfo {author} {\bibfnamefont {I.}~\bibnamefont
  {Vobornik}}, \bibinfo {author} {\bibfnamefont {B.}~\bibnamefont {Ghosh}},
  \bibinfo {author} {\bibfnamefont {A.}~\bibnamefont {Agarwal}}, \bibinfo
  {author} {\bibfnamefont {H.}~\bibnamefont {Xing}}, \bibinfo {author}
  {\bibfnamefont {X.}~\bibnamefont {Chen}}, \bibinfo {author} {\bibfnamefont
  {A.}~\bibnamefont {Politano}},\ and\ \bibinfo {author} {\bibfnamefont
  {W.}~\bibnamefont {Lu}},\ }\bibfield  {title} {\bibinfo {title}
  {High-frequency rectifiers based on {type-II Dirac} fermions},\ }\href
  {https://doi.org/10.1038/s41467-021-21906-w} {\bibfield  {journal} {\bibinfo
  {journal} {Nature Communications}\ }\textbf {\bibinfo {volume} {12}},\
  \bibinfo {pages} {1584} (\bibinfo {year} {2021})}\BibitemShut {NoStop}%
\bibitem [{\citenamefont {Zhang}\ \emph {et~al.}()\citenamefont {Zhang},
  \citenamefont {Guo}, \citenamefont {Kuo}, \citenamefont {Xu}, \citenamefont
  {Zhang}, \citenamefont {Ghosh}, \citenamefont {De~Santis}, \citenamefont
  {Boukhvalov}, \citenamefont {Vobornik}, \citenamefont {Paolucci},
  \citenamefont {Lue}, \citenamefont {Xing}, \citenamefont {Agarwal},
  \citenamefont {Wang},\ and\ \citenamefont {Politano}}]{NiTe2_THz_2}%
  \BibitemOpen
  \bibfield  {author} {\bibinfo {author} {\bibfnamefont {L.}~\bibnamefont
  {Zhang}}, \bibinfo {author} {\bibfnamefont {C.}~\bibnamefont {Guo}}, \bibinfo
  {author} {\bibfnamefont {C.-N.}\ \bibnamefont {Kuo}}, \bibinfo {author}
  {\bibfnamefont {H.}~\bibnamefont {Xu}}, \bibinfo {author} {\bibfnamefont
  {K.}~\bibnamefont {Zhang}}, \bibinfo {author} {\bibfnamefont
  {B.}~\bibnamefont {Ghosh}}, \bibinfo {author} {\bibfnamefont
  {J.}~\bibnamefont {De~Santis}}, \bibinfo {author} {\bibfnamefont {D.~W.}\
  \bibnamefont {Boukhvalov}}, \bibinfo {author} {\bibfnamefont
  {I.}~\bibnamefont {Vobornik}}, \bibinfo {author} {\bibfnamefont
  {V.}~\bibnamefont {Paolucci}}, \bibinfo {author} {\bibfnamefont {C.~S.}\
  \bibnamefont {Lue}}, \bibinfo {author} {\bibfnamefont {H.}~\bibnamefont
  {Xing}}, \bibinfo {author} {\bibfnamefont {A.}~\bibnamefont {Agarwal}},
  \bibinfo {author} {\bibfnamefont {L.}~\bibnamefont {Wang}},\ and\ \bibinfo
  {author} {\bibfnamefont {A.}~\bibnamefont {Politano}},\ }\bibfield  {title}
  {\bibinfo {title} {Terahertz photodetection with {Type-II Dirac} fermions in
  transition-metal ditellurides and their heterostructures},\ }\href
  {https://doi.org/https://doi.org/10.1002/pssr.202100212} {\bibfield
  {journal} {\bibinfo  {journal} {physica status solidi (RRL) – Rapid
  Research Letters}\ }\textbf {\bibinfo {volume} {n/a}},\ \bibinfo {pages}
  {2100212}}\BibitemShut {NoStop}%
\bibitem [{\citenamefont {Sirota}\ \emph {et~al.}(2018)\citenamefont {Sirota},
  \citenamefont {Glavin}, \citenamefont {Krylyuk}, \citenamefont {Davydov},\
  and\ \citenamefont {Voevodin}}]{Exf_MoTe2}%
  \BibitemOpen
  \bibfield  {author} {\bibinfo {author} {\bibfnamefont {B.}~\bibnamefont
  {Sirota}}, \bibinfo {author} {\bibfnamefont {N.}~\bibnamefont {Glavin}},
  \bibinfo {author} {\bibfnamefont {S.}~\bibnamefont {Krylyuk}}, \bibinfo
  {author} {\bibfnamefont {A.~V.}\ \bibnamefont {Davydov}},\ and\ \bibinfo
  {author} {\bibfnamefont {A.~A.}\ \bibnamefont {Voevodin}},\ }\bibfield
  {title} {\bibinfo {title} {Hexagonal {MoTe$_{2}$} with amorphous bn
  passivation layer for improved oxidation resistance and endurance of {2D}
  field effect transistors},\ }\href
  {https://doi.org/10.1038/s41598-018-26751-4} {\bibfield  {journal} {\bibinfo
  {journal} {Scientific Reports}\ }\textbf {\bibinfo {volume} {8}},\ \bibinfo
  {pages} {8668} (\bibinfo {year} {2018})}\BibitemShut {NoStop}%
\bibitem [{\citenamefont {Gołasa}\ \emph {et~al.}(2013)\citenamefont
  {Gołasa}, \citenamefont {Grzeszczyk}, \citenamefont {Korona}, \citenamefont
  {Bożek}, \citenamefont {Binder}, \citenamefont {Szczytko}, \citenamefont
  {Wysmołek},\ and\ \citenamefont {Babiński}}]{Golasa_MoS2}%
  \BibitemOpen
  \bibfield  {author} {\bibinfo {author} {\bibfnamefont {K.}~\bibnamefont
  {Gołasa}}, \bibinfo {author} {\bibfnamefont {M.}~\bibnamefont {Grzeszczyk}},
  \bibinfo {author} {\bibfnamefont {K.}~\bibnamefont {Korona}}, \bibinfo
  {author} {\bibfnamefont {R.}~\bibnamefont {Bożek}}, \bibinfo {author}
  {\bibfnamefont {J.}~\bibnamefont {Binder}}, \bibinfo {author} {\bibfnamefont
  {J.}~\bibnamefont {Szczytko}}, \bibinfo {author} {\bibfnamefont
  {A.}~\bibnamefont {Wysmołek}},\ and\ \bibinfo {author} {\bibfnamefont
  {A.}~\bibnamefont {Babiński}},\ }\bibfield  {title} {\bibinfo {title}
  {Optical properties of molybdenum disulfide {MoS$_{2}$}},\ }\href
  {https://doi.org/http://dx.doi.org/10.12693/APhysPolA.124.849} {\bibfield
  {journal} {\bibinfo  {journal} {Acta Physica Polonica A}\ }\textbf {\bibinfo
  {volume} {124}},\ \bibinfo {pages} {849} (\bibinfo {year}
  {2013})}\BibitemShut {NoStop}%
\bibitem [{\citenamefont {Pacuski}\ \emph {et~al.}(2020)\citenamefont
  {Pacuski}, \citenamefont {Grzeszczyk}, \citenamefont {Nogajewski},
  \citenamefont {Bogucki}, \citenamefont {Oreszczuk}, \citenamefont {Kucharek},
  \citenamefont {Połczyńska}, \citenamefont {Seredyński}, \citenamefont
  {Rodek}, \citenamefont {Bożek}, \citenamefont {Taniguchi}, \citenamefont
  {Watanabe}, \citenamefont {Kret}, \citenamefont {Sadowski}, \citenamefont
  {Kazimierczuk}, \citenamefont {Potemski},\ and\ \citenamefont
  {Kossacki}}]{MoSe2_nasze}%
  \BibitemOpen
  \bibfield  {author} {\bibinfo {author} {\bibfnamefont {W.}~\bibnamefont
  {Pacuski}}, \bibinfo {author} {\bibfnamefont {M.}~\bibnamefont {Grzeszczyk}},
  \bibinfo {author} {\bibfnamefont {K.}~\bibnamefont {Nogajewski}}, \bibinfo
  {author} {\bibfnamefont {A.}~\bibnamefont {Bogucki}}, \bibinfo {author}
  {\bibfnamefont {K.}~\bibnamefont {Oreszczuk}}, \bibinfo {author}
  {\bibfnamefont {J.}~\bibnamefont {Kucharek}}, \bibinfo {author}
  {\bibfnamefont {K.~E.}\ \bibnamefont {Połczyńska}}, \bibinfo {author}
  {\bibfnamefont {B.}~\bibnamefont {Seredyński}}, \bibinfo {author}
  {\bibfnamefont {A.}~\bibnamefont {Rodek}}, \bibinfo {author} {\bibfnamefont
  {R.}~\bibnamefont {Bożek}}, \bibinfo {author} {\bibfnamefont
  {T.}~\bibnamefont {Taniguchi}}, \bibinfo {author} {\bibfnamefont
  {K.}~\bibnamefont {Watanabe}}, \bibinfo {author} {\bibfnamefont
  {S.}~\bibnamefont {Kret}}, \bibinfo {author} {\bibfnamefont {J.}~\bibnamefont
  {Sadowski}}, \bibinfo {author} {\bibfnamefont {T.}~\bibnamefont
  {Kazimierczuk}}, \bibinfo {author} {\bibfnamefont {M.}~\bibnamefont
  {Potemski}},\ and\ \bibinfo {author} {\bibfnamefont {P.}~\bibnamefont
  {Kossacki}},\ }\bibfield  {title} {\bibinfo {title} {Narrow excitonic lines
  and large-scale homogeneity of transition-metal dichalcogenide monolayers
  grown by molecular beam epitaxy on hexagonal boron nitride},\ }\href
  {https://doi.org/10.1021/acs.nanolett.9b04998} {\bibfield  {journal}
  {\bibinfo  {journal} {Nano Letters}\ }\textbf {\bibinfo {volume} {20}},\
  \bibinfo {pages} {3058} (\bibinfo {year} {2020})}\BibitemShut {NoStop}%
\bibitem [{\citenamefont {Cai}\ \emph {et~al.}(2018)\citenamefont {Cai},
  \citenamefont {Liu}, \citenamefont {Zou},\ and\ \citenamefont
  {Cheng}}]{CVD_2D}%
  \BibitemOpen
  \bibfield  {author} {\bibinfo {author} {\bibfnamefont {Z.}~\bibnamefont
  {Cai}}, \bibinfo {author} {\bibfnamefont {B.}~\bibnamefont {Liu}}, \bibinfo
  {author} {\bibfnamefont {X.}~\bibnamefont {Zou}},\ and\ \bibinfo {author}
  {\bibfnamefont {H.-M.}\ \bibnamefont {Cheng}},\ }\bibfield  {title} {\bibinfo
  {title} {Chemical vapor deposition growth and applications of two-dimensional
  materials and their heterostructures},\ }\href
  {https://doi.org/10.1021/acs.chemrev.7b00536} {\bibfield  {journal} {\bibinfo
   {journal} {Chemical Reviews}\ }\textbf {\bibinfo {volume} {118}},\ \bibinfo
  {pages} {6091} (\bibinfo {year} {2018})}\BibitemShut {NoStop}%
\bibitem [{\citenamefont {Choi}\ \emph {et~al.}(2021)\citenamefont {Choi},
  \citenamefont {Kim}, \citenamefont {Song}, \citenamefont {Kim}, \citenamefont
  {Han}, \citenamefont {Nguyen}, \citenamefont {Ko}, \citenamefont {Boandoh},
  \citenamefont {Choi}, \citenamefont {Oh}, \citenamefont {Cho}, \citenamefont
  {Jin}, \citenamefont {Won}, \citenamefont {Lee}, \citenamefont {Yun},
  \citenamefont {Shin}, \citenamefont {Jeong}, \citenamefont {Kim},
  \citenamefont {Han}, \citenamefont {Lee}, \citenamefont {Kim},\ and\
  \citenamefont {Kim}}]{CVD_gold}%
  \BibitemOpen
  \bibfield  {author} {\bibinfo {author} {\bibfnamefont {S.~H.}\ \bibnamefont
  {Choi}}, \bibinfo {author} {\bibfnamefont {H.-J.}\ \bibnamefont {Kim}},
  \bibinfo {author} {\bibfnamefont {B.}~\bibnamefont {Song}}, \bibinfo {author}
  {\bibfnamefont {Y.~I.}\ \bibnamefont {Kim}}, \bibinfo {author} {\bibfnamefont
  {G.}~\bibnamefont {Han}}, \bibinfo {author} {\bibfnamefont {H.~T.~T.}\
  \bibnamefont {Nguyen}}, \bibinfo {author} {\bibfnamefont {H.}~\bibnamefont
  {Ko}}, \bibinfo {author} {\bibfnamefont {S.}~\bibnamefont {Boandoh}},
  \bibinfo {author} {\bibfnamefont {J.~H.}\ \bibnamefont {Choi}}, \bibinfo
  {author} {\bibfnamefont {C.~S.}\ \bibnamefont {Oh}}, \bibinfo {author}
  {\bibfnamefont {H.~J.}\ \bibnamefont {Cho}}, \bibinfo {author} {\bibfnamefont
  {J.~W.}\ \bibnamefont {Jin}}, \bibinfo {author} {\bibfnamefont {Y.~S.}\
  \bibnamefont {Won}}, \bibinfo {author} {\bibfnamefont {B.~H.}\ \bibnamefont
  {Lee}}, \bibinfo {author} {\bibfnamefont {S.~J.}\ \bibnamefont {Yun}},
  \bibinfo {author} {\bibfnamefont {B.~G.}\ \bibnamefont {Shin}}, \bibinfo
  {author} {\bibfnamefont {H.~Y.}\ \bibnamefont {Jeong}}, \bibinfo {author}
  {\bibfnamefont {Y.-M.}\ \bibnamefont {Kim}}, \bibinfo {author} {\bibfnamefont
  {Y.-K.}\ \bibnamefont {Han}}, \bibinfo {author} {\bibfnamefont {Y.~H.}\
  \bibnamefont {Lee}}, \bibinfo {author} {\bibfnamefont {S.~M.}\ \bibnamefont
  {Kim}},\ and\ \bibinfo {author} {\bibfnamefont {K.~K.}\ \bibnamefont {Kim}},\
  }\bibfield  {title} {\bibinfo {title} {Epitaxial single-crystal growth of
  transition metal dichalcogenide monolayers via the atomic sawtooth {Au}
  surface},\ }\href {https://doi.org/https://doi.org/10.1002/adma.202006601}
  {\bibfield  {journal} {\bibinfo  {journal} {Advanced Materials}\ }\textbf
  {\bibinfo {volume} {33}},\ \bibinfo {pages} {2006601} (\bibinfo {year}
  {2021})}\BibitemShut {NoStop}%
\bibitem [{\citenamefont {Poh}\ \emph {et~al.}(2017)\citenamefont {Poh},
  \citenamefont {Tan}, \citenamefont {Zhao}, \citenamefont {Chen},
  \citenamefont {Abdelwahab}, \citenamefont {Fu}, \citenamefont {Xu},
  \citenamefont {Bao}, \citenamefont {Zhou},\ and\ \citenamefont
  {Loh}}]{Large_area_MoSe2}%
  \BibitemOpen
  \bibfield  {author} {\bibinfo {author} {\bibfnamefont {S.~M.}\ \bibnamefont
  {Poh}}, \bibinfo {author} {\bibfnamefont {S.~J.~R.}\ \bibnamefont {Tan}},
  \bibinfo {author} {\bibfnamefont {X.}~\bibnamefont {Zhao}}, \bibinfo {author}
  {\bibfnamefont {Z.}~\bibnamefont {Chen}}, \bibinfo {author} {\bibfnamefont
  {I.}~\bibnamefont {Abdelwahab}}, \bibinfo {author} {\bibfnamefont
  {D.}~\bibnamefont {Fu}}, \bibinfo {author} {\bibfnamefont {H.}~\bibnamefont
  {Xu}}, \bibinfo {author} {\bibfnamefont {Y.}~\bibnamefont {Bao}}, \bibinfo
  {author} {\bibfnamefont {W.}~\bibnamefont {Zhou}},\ and\ \bibinfo {author}
  {\bibfnamefont {K.~P.}\ \bibnamefont {Loh}},\ }\bibfield  {title} {\bibinfo
  {title} {Large area synthesis of {1D-MoSe$_{2}$} using molecular beam
  epitaxy},\ }\href {https://doi.org/10.1002/adma.201605641} {\bibfield
  {journal} {\bibinfo  {journal} {Advanced Materials}\ }\textbf {\bibinfo
  {volume} {29}},\ \bibinfo {pages} {1605641} (\bibinfo {year}
  {2017})}\BibitemShut {NoStop}%
\bibitem [{\citenamefont {Vishwanath}\ \emph {et~al.}(2016)\citenamefont
  {Vishwanath}, \citenamefont {Liu}, \citenamefont {Rouvimov}, \citenamefont
  {Basile}, \citenamefont {Lu}, \citenamefont {Azcatl}, \citenamefont {Magno},
  \citenamefont {Wallace}, \citenamefont {Kim}, \citenamefont {Idrobo},
  \citenamefont {Furdyna}, \citenamefont {Jena},\ and\ \citenamefont
  {Xing}}]{Furdyna_przeglad}%
  \BibitemOpen
  \bibfield  {author} {\bibinfo {author} {\bibfnamefont {S.}~\bibnamefont
  {Vishwanath}}, \bibinfo {author} {\bibfnamefont {X.}~\bibnamefont {Liu}},
  \bibinfo {author} {\bibfnamefont {S.}~\bibnamefont {Rouvimov}}, \bibinfo
  {author} {\bibfnamefont {L.}~\bibnamefont {Basile}}, \bibinfo {author}
  {\bibfnamefont {N.}~\bibnamefont {Lu}}, \bibinfo {author} {\bibfnamefont
  {A.}~\bibnamefont {Azcatl}}, \bibinfo {author} {\bibfnamefont
  {K.}~\bibnamefont {Magno}}, \bibinfo {author} {\bibfnamefont {R.~M.}\
  \bibnamefont {Wallace}}, \bibinfo {author} {\bibfnamefont {M.}~\bibnamefont
  {Kim}}, \bibinfo {author} {\bibfnamefont {J.-C.}\ \bibnamefont {Idrobo}},
  \bibinfo {author} {\bibfnamefont {J.~K.}\ \bibnamefont {Furdyna}}, \bibinfo
  {author} {\bibfnamefont {D.}~\bibnamefont {Jena}},\ and\ \bibinfo {author}
  {\bibfnamefont {H.~G.}\ \bibnamefont {Xing}},\ }\bibfield  {title} {\bibinfo
  {title} {Controllable growth of layered selenide and telluride
  heterostructures and superlattices using molecular beam epitaxy},\ }\href
  {https://doi.org/10.1557/jmr.2015.374} {\bibfield  {journal} {\bibinfo
  {journal} {Journal of Materials Research}\ }\textbf {\bibinfo {volume}
  {31}},\ \bibinfo {pages} {900} (\bibinfo {year} {2016})}\BibitemShut
  {NoStop}%
\bibitem [{\citenamefont {Ogorzałek}\ \emph {et~al.}(2020)\citenamefont
  {Ogorzałek}, \citenamefont {Seredyński}, \citenamefont {Kret},
  \citenamefont {Kwiatkowski}, \citenamefont {Korona}, \citenamefont
  {Grzeszczyk}, \citenamefont {Mierzejewski}, \citenamefont {Wasik},
  \citenamefont {Pacuski}, \citenamefont {Sadowski},\ and\ \citenamefont
  {Gryglas-Borysiewicz}}]{MoTe2_nasze}%
  \BibitemOpen
  \bibfield  {author} {\bibinfo {author} {\bibfnamefont {Z.}~\bibnamefont
  {Ogorzałek}}, \bibinfo {author} {\bibfnamefont {B.}~\bibnamefont
  {Seredyński}}, \bibinfo {author} {\bibfnamefont {S.}~\bibnamefont {Kret}},
  \bibinfo {author} {\bibfnamefont {A.}~\bibnamefont {Kwiatkowski}}, \bibinfo
  {author} {\bibfnamefont {K.~P.}\ \bibnamefont {Korona}}, \bibinfo {author}
  {\bibfnamefont {M.}~\bibnamefont {Grzeszczyk}}, \bibinfo {author}
  {\bibfnamefont {J.}~\bibnamefont {Mierzejewski}}, \bibinfo {author}
  {\bibfnamefont {D.}~\bibnamefont {Wasik}}, \bibinfo {author} {\bibfnamefont
  {W.}~\bibnamefont {Pacuski}}, \bibinfo {author} {\bibfnamefont
  {J.}~\bibnamefont {Sadowski}},\ and\ \bibinfo {author} {\bibfnamefont
  {M.}~\bibnamefont {Gryglas-Borysiewicz}},\ }\bibfield  {title} {\bibinfo
  {title} {Charge transport in mbe-grown {2H-MoTe$_{2}$} bilayers with enhanced
  stability provided by an alox capping layer},\ }\href
  {https://doi.org/10.1039/D0NR03148H} {\bibfield  {journal} {\bibinfo
  {journal} {Nanoscale}\ }\textbf {\bibinfo {volume} {12}},\ \bibinfo {pages}
  {16535} (\bibinfo {year} {2020})}\BibitemShut {NoStop}%
\bibitem [{\citenamefont {Lasek}\ \emph {et~al.}(2020)\citenamefont {Lasek},
  \citenamefont {Coelho}, \citenamefont {Zberecki}, \citenamefont {Xin},
  \citenamefont {Kolekar}, \citenamefont {Li},\ and\ \citenamefont
  {Batzill}}]{MBE_TMDC_tellurides}%
  \BibitemOpen
  \bibfield  {author} {\bibinfo {author} {\bibfnamefont {K.}~\bibnamefont
  {Lasek}}, \bibinfo {author} {\bibfnamefont {P.~M.}\ \bibnamefont {Coelho}},
  \bibinfo {author} {\bibfnamefont {K.}~\bibnamefont {Zberecki}}, \bibinfo
  {author} {\bibfnamefont {Y.}~\bibnamefont {Xin}}, \bibinfo {author}
  {\bibfnamefont {S.~K.}\ \bibnamefont {Kolekar}}, \bibinfo {author}
  {\bibfnamefont {J.}~\bibnamefont {Li}},\ and\ \bibinfo {author}
  {\bibfnamefont {M.}~\bibnamefont {Batzill}},\ }\bibfield  {title} {\bibinfo
  {title} {Molecular beam epitaxy of transition metal {(Ti-, V-, and Cr-)}
  tellurides: From monolayer ditellurides to multilayer self-intercalation
  compounds},\ }\href {https://doi.org/10.1021/acsnano.0c02712} {\bibfield
  {journal} {\bibinfo  {journal} {ACS Nano}\ }\textbf {\bibinfo {volume}
  {14}},\ \bibinfo {pages} {8473} (\bibinfo {year} {2020})},\ \Eprint
  {https://arxiv.org/abs/https://doi.org/10.1021/acsnano.0c02712}
  {https://doi.org/10.1021/acsnano.0c02712} \BibitemShut {NoStop}%
\bibitem [{\citenamefont {Rajan}\ \emph {et~al.}(2020)\citenamefont {Rajan},
  \citenamefont {Underwood}, \citenamefont {Mazzola},\ and\ \citenamefont
  {King}}]{MBE_TMDC_selenides}%
  \BibitemOpen
  \bibfield  {author} {\bibinfo {author} {\bibfnamefont {A.}~\bibnamefont
  {Rajan}}, \bibinfo {author} {\bibfnamefont {K.}~\bibnamefont {Underwood}},
  \bibinfo {author} {\bibfnamefont {F.}~\bibnamefont {Mazzola}},\ and\ \bibinfo
  {author} {\bibfnamefont {P.~D.~C.}\ \bibnamefont {King}},\ }\bibfield
  {title} {\bibinfo {title} {Morphology control of epitaxial monolayer
  transition metal dichalcogenides},\ }\href
  {https://doi.org/10.1103/PhysRevMaterials.4.014003} {\bibfield  {journal}
  {\bibinfo  {journal} {Phys. Rev. Materials}\ }\textbf {\bibinfo {volume}
  {4}},\ \bibinfo {pages} {014003} (\bibinfo {year} {2020})}\BibitemShut
  {NoStop}%
\bibitem [{\citenamefont {Ohtake}\ and\ \citenamefont
  {Sakuma}(2020)}]{MoSe2_orientation}%
  \BibitemOpen
  \bibfield  {author} {\bibinfo {author} {\bibfnamefont {A.}~\bibnamefont
  {Ohtake}}\ and\ \bibinfo {author} {\bibfnamefont {Y.}~\bibnamefont
  {Sakuma}},\ }\bibfield  {title} {\bibinfo {title} {Effect of substrate
  orientation on {MoSe$_{2}$/GaAs} heteroepitaxy},\ }\href
  {https://doi.org/10.1021/acs.jpcc.9b11278} {\bibfield  {journal} {\bibinfo
  {journal} {The Journal of Physical Chemistry C}\ }\textbf {\bibinfo {volume}
  {124}},\ \bibinfo {pages} {5196} (\bibinfo {year} {2020})},\ \Eprint
  {https://arxiv.org/abs/https://doi.org/10.1021/acs.jpcc.9b11278}
  {https://doi.org/10.1021/acs.jpcc.9b11278} \BibitemShut {NoStop}%
\bibitem [{\citenamefont {Monteiro}\ \emph
  {et~al.}(2017{\natexlab{a}})\citenamefont {Monteiro}, \citenamefont
  {Marciniak}, \citenamefont {Jurelo}, \citenamefont {Siqueira}, \citenamefont
  {Dias},\ and\ \citenamefont {{Pimentel Júnior}}}]{Bulk_1}%
  \BibitemOpen
  \bibfield  {author} {\bibinfo {author} {\bibfnamefont {J.~F. H.~L.}\
  \bibnamefont {Monteiro}}, \bibinfo {author} {\bibfnamefont {M.~B.}\
  \bibnamefont {Marciniak}}, \bibinfo {author} {\bibfnamefont {A.~R.}\
  \bibnamefont {Jurelo}}, \bibinfo {author} {\bibfnamefont {E.~C.}\
  \bibnamefont {Siqueira}}, \bibinfo {author} {\bibfnamefont {F.~T.}\
  \bibnamefont {Dias}},\ and\ \bibinfo {author} {\bibfnamefont {J.~L.}\
  \bibnamefont {{Pimentel Júnior}}},\ }\bibfield  {title} {\bibinfo {title}
  {Synthesis and microstructure of {NiTe$_{2}$}},\ }\href
  {https://doi.org/https://doi.org/10.1016/j.jcrysgro.2017.08.030} {\bibfield
  {journal} {\bibinfo  {journal} {Journal of Crystal Growth}\ }\textbf
  {\bibinfo {volume} {478}},\ \bibinfo {pages} {129 } (\bibinfo {year}
  {2017}{\natexlab{a}})}\BibitemShut {NoStop}%
\bibitem [{\citenamefont {Liu}\ \emph {et~al.}(2009)\citenamefont {Liu},
  \citenamefont {Hu}, \citenamefont {Chai}, \citenamefont {Li}, \citenamefont
  {Gu},\ and\ \citenamefont {Qian}}]{Bulk_2}%
  \BibitemOpen
  \bibfield  {author} {\bibinfo {author} {\bibfnamefont {X.}~\bibnamefont
  {Liu}}, \bibinfo {author} {\bibfnamefont {R.}~\bibnamefont {Hu}}, \bibinfo
  {author} {\bibfnamefont {L.}~\bibnamefont {Chai}}, \bibinfo {author}
  {\bibfnamefont {H.}~\bibnamefont {Li}}, \bibinfo {author} {\bibfnamefont
  {J.}~\bibnamefont {Gu}},\ and\ \bibinfo {author} {\bibfnamefont
  {Y.}~\bibnamefont {Qian}},\ }\bibfield  {title} {\bibinfo {title} {Synthesis
  and characterization of hexagonal {NiTe$_{2}$} nanoplates},\ }\href
  {https://doi.org/https://doi.org/10.1166/jnn.2009.dk14} {\bibfield  {journal}
  {\bibinfo  {journal} {Journal of Nanoscience and Nanotechnology}\ }\textbf
  {\bibinfo {volume} {9}},\ \bibinfo {pages} {2715} (\bibinfo {year}
  {2009})}\BibitemShut {NoStop}%
\bibitem [{\citenamefont {Liu}\ \emph {et~al.}(2019)\citenamefont {Liu},
  \citenamefont {Fei}, \citenamefont {Chen}, \citenamefont {Bo}, \citenamefont
  {Wei}, \citenamefont {Zhang}, \citenamefont {Zhang}, \citenamefont {Xie},
  \citenamefont {Naveed}, \citenamefont {Wan}, \citenamefont {Song},\ and\
  \citenamefont {Wang}}]{Bulk_3}%
  \BibitemOpen
  \bibfield  {author} {\bibinfo {author} {\bibfnamefont {Q.}~\bibnamefont
  {Liu}}, \bibinfo {author} {\bibfnamefont {F.}~\bibnamefont {Fei}}, \bibinfo
  {author} {\bibfnamefont {B.}~\bibnamefont {Chen}}, \bibinfo {author}
  {\bibfnamefont {X.}~\bibnamefont {Bo}}, \bibinfo {author} {\bibfnamefont
  {B.}~\bibnamefont {Wei}}, \bibinfo {author} {\bibfnamefont {S.}~\bibnamefont
  {Zhang}}, \bibinfo {author} {\bibfnamefont {M.}~\bibnamefont {Zhang}},
  \bibinfo {author} {\bibfnamefont {F.}~\bibnamefont {Xie}}, \bibinfo {author}
  {\bibfnamefont {M.}~\bibnamefont {Naveed}}, \bibinfo {author} {\bibfnamefont
  {X.}~\bibnamefont {Wan}}, \bibinfo {author} {\bibfnamefont {F.}~\bibnamefont
  {Song}},\ and\ \bibinfo {author} {\bibfnamefont {B.}~\bibnamefont {Wang}},\
  }\bibfield  {title} {\bibinfo {title} {Nontopological origin of the planar
  {Hall} effect in the type-{II Dirac} semimetal {${\mathrm{NiTe}}_{2}$}},\
  }\href {https://doi.org/10.1103/PhysRevB.99.155119} {\bibfield  {journal}
  {\bibinfo  {journal} {Phys. Rev. B}\ }\textbf {\bibinfo {volume} {99}},\
  \bibinfo {pages} {155119} (\bibinfo {year} {2019})}\BibitemShut {NoStop}%
\bibitem [{\citenamefont {Ghosh}\ \emph {et~al.}(2019)\citenamefont {Ghosh},
  \citenamefont {Mondal}, \citenamefont {Kuo}, \citenamefont {Lue},
  \citenamefont {Nayak}, \citenamefont {Fujii}, \citenamefont {Vobornik},
  \citenamefont {Politano},\ and\ \citenamefont {Agarwal}}]{Bulk_4}%
  \BibitemOpen
  \bibfield  {author} {\bibinfo {author} {\bibfnamefont {B.}~\bibnamefont
  {Ghosh}}, \bibinfo {author} {\bibfnamefont {D.}~\bibnamefont {Mondal}},
  \bibinfo {author} {\bibfnamefont {C.-N.}\ \bibnamefont {Kuo}}, \bibinfo
  {author} {\bibfnamefont {C.~S.}\ \bibnamefont {Lue}}, \bibinfo {author}
  {\bibfnamefont {J.}~\bibnamefont {Nayak}}, \bibinfo {author} {\bibfnamefont
  {J.}~\bibnamefont {Fujii}}, \bibinfo {author} {\bibfnamefont
  {I.}~\bibnamefont {Vobornik}}, \bibinfo {author} {\bibfnamefont
  {A.}~\bibnamefont {Politano}},\ and\ \bibinfo {author} {\bibfnamefont
  {A.}~\bibnamefont {Agarwal}},\ }\bibfield  {title} {\bibinfo {title}
  {Observation of bulk states and spin-polarized topological surface states in
  transition metal dichalcogenide {Dirac} semimetal candidate
  {${\mathrm{NiTe}}_{2}$}},\ }\href
  {https://doi.org/10.1103/PhysRevB.100.195134} {\bibfield  {journal} {\bibinfo
   {journal} {Phys. Rev. B}\ }\textbf {\bibinfo {volume} {100}},\ \bibinfo
  {pages} {195134} (\bibinfo {year} {2019})}\BibitemShut {NoStop}%
\bibitem [{\citenamefont {Dulal}\ \emph {et~al.}(2019)\citenamefont {Dulal},
  \citenamefont {Dahal}, \citenamefont {Forbes}, \citenamefont {Bhattarai},
  \citenamefont {Pegg},\ and\ \citenamefont {Philip}}]{Bulk_5}%
  \BibitemOpen
  \bibfield  {author} {\bibinfo {author} {\bibfnamefont {R.~P.}\ \bibnamefont
  {Dulal}}, \bibinfo {author} {\bibfnamefont {B.~R.}\ \bibnamefont {Dahal}},
  \bibinfo {author} {\bibfnamefont {A.}~\bibnamefont {Forbes}}, \bibinfo
  {author} {\bibfnamefont {N.}~\bibnamefont {Bhattarai}}, \bibinfo {author}
  {\bibfnamefont {I.~L.}\ \bibnamefont {Pegg}},\ and\ \bibinfo {author}
  {\bibfnamefont {J.}~\bibnamefont {Philip}},\ }\bibfield  {title} {\bibinfo
  {title} {Nanostructures of type-{II} topological {Dirac} semimetal
  {NiTe$_{2}$}},\ }\href {https://doi.org/10.1116/1.5111331} {\bibfield
  {journal} {\bibinfo  {journal} {Journal of Vacuum Science \& Technology B}\
  }\textbf {\bibinfo {volume} {37}},\ \bibinfo {pages} {042903} (\bibinfo
  {year} {2019})}\BibitemShut {NoStop}%
\bibitem [{\citenamefont {Zhao}\ \emph {et~al.}(2018)\citenamefont {Zhao},
  \citenamefont {Dang}, \citenamefont {Liu}, \citenamefont {Li}, \citenamefont
  {Li}, \citenamefont {Luo}, \citenamefont {Zhang}, \citenamefont {Wu},
  \citenamefont {Ma}, \citenamefont {Sun}, \citenamefont {Huang}, \citenamefont
  {Duan},\ and\ \citenamefont {Duan}}]{CVD_JACS}%
  \BibitemOpen
  \bibfield  {author} {\bibinfo {author} {\bibfnamefont {B.}~\bibnamefont
  {Zhao}}, \bibinfo {author} {\bibfnamefont {W.}~\bibnamefont {Dang}}, \bibinfo
  {author} {\bibfnamefont {Y.}~\bibnamefont {Liu}}, \bibinfo {author}
  {\bibfnamefont {B.}~\bibnamefont {Li}}, \bibinfo {author} {\bibfnamefont
  {J.}~\bibnamefont {Li}}, \bibinfo {author} {\bibfnamefont {J.}~\bibnamefont
  {Luo}}, \bibinfo {author} {\bibfnamefont {Z.}~\bibnamefont {Zhang}}, \bibinfo
  {author} {\bibfnamefont {R.}~\bibnamefont {Wu}}, \bibinfo {author}
  {\bibfnamefont {H.}~\bibnamefont {Ma}}, \bibinfo {author} {\bibfnamefont
  {G.}~\bibnamefont {Sun}}, \bibinfo {author} {\bibfnamefont {Y.}~\bibnamefont
  {Huang}}, \bibinfo {author} {\bibfnamefont {X.}~\bibnamefont {Duan}},\ and\
  \bibinfo {author} {\bibfnamefont {X.}~\bibnamefont {Duan}},\ }\bibfield
  {title} {\bibinfo {title} {Synthetic control of two-dimensional {NiTe$_{2}$}
  single crystals with highly uniform thickness distributions},\ }\href
  {https://doi.org/10.1021/jacs.8b08124} {\bibfield  {journal} {\bibinfo
  {journal} {Journal of the American Chemical Society}\ }\textbf {\bibinfo
  {volume} {140}},\ \bibinfo {pages} {14217} (\bibinfo {year}
  {2018})}\BibitemShut {NoStop}%
\bibitem [{\citenamefont {Zhai}\ \emph {et~al.}(2020)\citenamefont {Zhai},
  \citenamefont {Xu}, \citenamefont {Peng}, \citenamefont {Jing}, \citenamefont
  {Zhang}, \citenamefont {Liu},\ and\ \citenamefont {Hu}}]{NiTe2_na_MoS2}%
  \BibitemOpen
  \bibfield  {author} {\bibinfo {author} {\bibfnamefont {X.}~\bibnamefont
  {Zhai}}, \bibinfo {author} {\bibfnamefont {X.}~\bibnamefont {Xu}}, \bibinfo
  {author} {\bibfnamefont {J.}~\bibnamefont {Peng}}, \bibinfo {author}
  {\bibfnamefont {F.}~\bibnamefont {Jing}}, \bibinfo {author} {\bibfnamefont
  {Q.}~\bibnamefont {Zhang}}, \bibinfo {author} {\bibfnamefont
  {H.}~\bibnamefont {Liu}},\ and\ \bibinfo {author} {\bibfnamefont
  {Z.}~\bibnamefont {Hu}},\ }\bibfield  {title} {\bibinfo {title} {Enhanced
  optoelectronic performance of cvd-grown metal–semiconductor
  {NiTe$_{2}$/MoS$_{2}$} heterostructures},\ }\href
  {https://doi.org/10.1021/acsami.0c02166} {\bibfield  {journal} {\bibinfo
  {journal} {ACS Applied Materials \& Interfaces}\ }\textbf {\bibinfo {volume}
  {12}},\ \bibinfo {pages} {24093} (\bibinfo {year} {2020})}\BibitemShut
  {NoStop}%
\bibitem [{\citenamefont {Shi}\ \emph {et~al.}(2020{\natexlab{b}})\citenamefont
  {Shi}, \citenamefont {Huan}, \citenamefont {Xiao}, \citenamefont {Hong},
  \citenamefont {Zhao}, \citenamefont {Gao}, \citenamefont {Cui}, \citenamefont
  {Yang}, \citenamefont {Pennycook}, \citenamefont {Zhao},\ and\ \citenamefont
  {Zhang}}]{CVD_mica}%
  \BibitemOpen
  \bibfield  {author} {\bibinfo {author} {\bibfnamefont {J.}~\bibnamefont
  {Shi}}, \bibinfo {author} {\bibfnamefont {Y.}~\bibnamefont {Huan}}, \bibinfo
  {author} {\bibfnamefont {M.}~\bibnamefont {Xiao}}, \bibinfo {author}
  {\bibfnamefont {M.}~\bibnamefont {Hong}}, \bibinfo {author} {\bibfnamefont
  {X.}~\bibnamefont {Zhao}}, \bibinfo {author} {\bibfnamefont {Y.}~\bibnamefont
  {Gao}}, \bibinfo {author} {\bibfnamefont {F.}~\bibnamefont {Cui}}, \bibinfo
  {author} {\bibfnamefont {P.}~\bibnamefont {Yang}}, \bibinfo {author}
  {\bibfnamefont {S.~J.}\ \bibnamefont {Pennycook}}, \bibinfo {author}
  {\bibfnamefont {J.}~\bibnamefont {Zhao}},\ and\ \bibinfo {author}
  {\bibfnamefont {Y.}~\bibnamefont {Zhang}},\ }\bibfield  {title} {\bibinfo
  {title} {Two-dimensional metallic {NiTe$_{2}$} with ultrahigh environmental
  stability, conductivity, and electrocatalytic activity},\ }\href
  {https://doi.org/10.1021/acsnano.0c03940} {\bibfield  {journal} {\bibinfo
  {journal} {ACS Nano}\ }\textbf {\bibinfo {volume} {14}},\ \bibinfo {pages}
  {9011} (\bibinfo {year} {2020}{\natexlab{b}})},\ \Eprint
  {https://arxiv.org/abs/https://doi.org/10.1021/acsnano.0c03940}
  {https://doi.org/10.1021/acsnano.0c03940} \BibitemShut {NoStop}%
\bibitem [{\citenamefont {Ross}\ \emph {et~al.}(1991)\citenamefont {Ross},
  \citenamefont {Rubin},\ and\ \citenamefont {Gustafson}}]{GaAs_lattice}%
  \BibitemOpen
  \bibfield  {author} {\bibinfo {author} {\bibfnamefont {J.~T.}\ \bibnamefont
  {Ross}}, \bibinfo {author} {\bibfnamefont {M.~D.}\ \bibnamefont {Rubin}},\
  and\ \bibinfo {author} {\bibfnamefont {T.~K.}\ \bibnamefont {Gustafson}},\
  }\bibfield  {title} {\bibinfo {title} {Single crystal wurtzite gan on (111)
  gaas with ain buffer layers grown by reactive magnetron sputter deposition},\
  }\href@noop {} {\bibfield  {journal} {\bibinfo  {journal} {Journal of
  Materials Research}\ }\textbf {\bibinfo {volume} {8}},\ \bibinfo {pages}
  {2613} (\bibinfo {year} {1991})}\BibitemShut {NoStop}%
\bibitem [{\citenamefont {Ohtake}\ \emph {et~al.}(2001)\citenamefont {Ohtake},
  \citenamefont {Nakamura}, \citenamefont {Komura}, \citenamefont {Hanada},
  \citenamefont {Yao}, \citenamefont {Kuramochi},\ and\ \citenamefont
  {Ozeki}}]{GaAs111_reconstruction}%
  \BibitemOpen
  \bibfield  {author} {\bibinfo {author} {\bibfnamefont {A.}~\bibnamefont
  {Ohtake}}, \bibinfo {author} {\bibfnamefont {J.}~\bibnamefont {Nakamura}},
  \bibinfo {author} {\bibfnamefont {T.}~\bibnamefont {Komura}}, \bibinfo
  {author} {\bibfnamefont {T.}~\bibnamefont {Hanada}}, \bibinfo {author}
  {\bibfnamefont {T.}~\bibnamefont {Yao}}, \bibinfo {author} {\bibfnamefont
  {H.}~\bibnamefont {Kuramochi}},\ and\ \bibinfo {author} {\bibfnamefont
  {M.}~\bibnamefont {Ozeki}},\ }\bibfield  {title} {\bibinfo {title} {Surface
  structures of
  $\mathrm{GaAs}\mathrm{{}111\mathrm{}}a,b\ensuremath{-}(2\ifmmode\times\else\texttimes\fi{}2)$},\
  }\href {https://doi.org/10.1103/PhysRevB.64.045318} {\bibfield  {journal}
  {\bibinfo  {journal} {Phys. Rev. B}\ }\textbf {\bibinfo {volume} {64}},\
  \bibinfo {pages} {045318} (\bibinfo {year} {2001})}\BibitemShut {NoStop}%
\bibitem [{baz({\natexlab{a}})}]{baza_XRD}%
  \BibitemOpen
  \href@noop {} {\bibfield  {journal} {\bibinfo  {journal} {© 2020
  International Centre for Diffraction Data. All rights reserved.}\ ,\ \bibinfo
  {pages} {card no. 00 008 0004}} ({\natexlab{a}})}\BibitemShut {NoStop}%
\bibitem [{baz({\natexlab{b}})}]{baza_XRD2}%
  \BibitemOpen
  \href@noop {} {\bibfield  {journal} {\bibinfo  {journal} {© 2020
  International Centre for Diffraction Data. All rights reserved.}\ ,\ \bibinfo
  {pages} {card no. 01}} ({\natexlab{b}})}\BibitemShut {NoStop}%
\bibitem [{\citenamefont {Monteiro}\ \emph
  {et~al.}(2017{\natexlab{b}})\citenamefont {Monteiro}, \citenamefont
  {Marciniak}, \citenamefont {Jurelo}, \citenamefont {Siqueira}, \citenamefont
  {Dias},\ and\ \citenamefont {{Pimentel Júnior}}}]{XRD_stala_c}%
  \BibitemOpen
  \bibfield  {author} {\bibinfo {author} {\bibfnamefont {J.~F. H.~L.}\
  \bibnamefont {Monteiro}}, \bibinfo {author} {\bibfnamefont {M.~B.}\
  \bibnamefont {Marciniak}}, \bibinfo {author} {\bibfnamefont {A.~R.}\
  \bibnamefont {Jurelo}}, \bibinfo {author} {\bibfnamefont {E.~C.}\
  \bibnamefont {Siqueira}}, \bibinfo {author} {\bibfnamefont {F.~T.}\
  \bibnamefont {Dias}},\ and\ \bibinfo {author} {\bibfnamefont {J.~L.}\
  \bibnamefont {{Pimentel Júnior}}},\ }\bibfield  {title} {\bibinfo {title}
  {Synthesis and microstructure of {$\mathrm{Ni}{\mathrm{Te}}_{2}$}},\ }\href
  {https://doi.org/https://doi.org/10.1016/j.jcrysgro.2017.08.030} {\bibfield
  {journal} {\bibinfo  {journal} {Journal of Crystal Growth}\ }\textbf
  {\bibinfo {volume} {478}},\ \bibinfo {pages} {129} (\bibinfo {year}
  {2017}{\natexlab{b}})}\BibitemShut {NoStop}%
\bibitem [{\citenamefont {Zheng}\ \emph {et~al.}(2020)\citenamefont {Zheng},
  \citenamefont {Sch\"onemann}, \citenamefont {Mozaffari}, \citenamefont
  {Chiu}, \citenamefont {Goraum}, \citenamefont {Aryal}, \citenamefont
  {Manousakis}, \citenamefont {Siegrist}, \citenamefont {Wei},\ and\
  \citenamefont {Balicas}}]{PRB2020}%
  \BibitemOpen
  \bibfield  {author} {\bibinfo {author} {\bibfnamefont {W.}~\bibnamefont
  {Zheng}}, \bibinfo {author} {\bibfnamefont {R.}~\bibnamefont {Sch\"onemann}},
  \bibinfo {author} {\bibfnamefont {S.}~\bibnamefont {Mozaffari}}, \bibinfo
  {author} {\bibfnamefont {Y.-C.}\ \bibnamefont {Chiu}}, \bibinfo {author}
  {\bibfnamefont {Z.~B.}\ \bibnamefont {Goraum}}, \bibinfo {author}
  {\bibfnamefont {N.}~\bibnamefont {Aryal}}, \bibinfo {author} {\bibfnamefont
  {E.}~\bibnamefont {Manousakis}}, \bibinfo {author} {\bibfnamefont {T.~M.}\
  \bibnamefont {Siegrist}}, \bibinfo {author} {\bibfnamefont {K.}~\bibnamefont
  {Wei}},\ and\ \bibinfo {author} {\bibfnamefont {L.}~\bibnamefont {Balicas}},\
  }\bibfield  {title} {\bibinfo {title} {Bulk fermi surfaces of the {Dirac}
  type-{II} semimetallic candidate {$\mathrm{Ni}{\mathrm{Te}}_{2}$}},\ }\href
  {https://doi.org/10.1103/PhysRevB.102.125103} {\bibfield  {journal} {\bibinfo
   {journal} {Phys. Rev. B}\ }\textbf {\bibinfo {volume} {102}},\ \bibinfo
  {pages} {125103} (\bibinfo {year} {2020})}\BibitemShut {NoStop}%
\bibitem [{\citenamefont {Jiang}\ \emph {et~al.}(2020)\citenamefont {Jiang},
  \citenamefont {Wei}, \citenamefont {Zhao}, \citenamefont {Wang},
  \citenamefont {Fang},\ and\ \citenamefont {Zhang}}]{SERS_Raman}%
  \BibitemOpen
  \bibfield  {author} {\bibinfo {author} {\bibfnamefont {C.}~\bibnamefont
  {Jiang}}, \bibinfo {author} {\bibfnamefont {Y.}~\bibnamefont {Wei}}, \bibinfo
  {author} {\bibfnamefont {P.}~\bibnamefont {Zhao}}, \bibinfo {author}
  {\bibfnamefont {P.}~\bibnamefont {Wang}}, \bibinfo {author} {\bibfnamefont
  {Y.}~\bibnamefont {Fang}},\ and\ \bibinfo {author} {\bibfnamefont
  {L.}~\bibnamefont {Zhang}},\ }\bibfield  {title} {\bibinfo {title}
  {Investigation of surface-enhanced raman spectroscopy on the substrates of
  telluride {2D} material},\ }\href
  {https://doi.org/10.1140/epjp/s13360-020-00688-y} {\bibfield  {journal}
  {\bibinfo  {journal} {The European Physical Journal Plus}\ }\textbf {\bibinfo
  {volume} {135}},\ \bibinfo {pages} {671} (\bibinfo {year}
  {2020})}\BibitemShut {NoStop}%
\end{thebibliography}

\end{document}